\documentclass[apj]{emulateapj}
\usepackage{graphicx}
\usepackage{epsfig}

\newcommand{\cm}{\textrm{\thinspace cm}}
\newcommand{\cts}{\textrm{\thinspace cts}}
\newcommand{\erg}{\textrm{\thinspace erg}}
\newcommand{\eV}{\textrm{\thinspace eV}}

\newcommand{\keV}{\textrm{\thinspace\,keV}}
\newcommand{\km}{\textrm{\thinspace km}}

\newcommand{\ks}{\textrm{\thinspace ks}}

\newcommand{\Mpc}{\textrm{\thinspace Mpc}}
\newcommand{\Msun}{\hbox{$\mathrm{\thinspace M_{\odot}}$}}
\newcommand{\pc}{\textrm{\thinspace pc}}
\newcommand{\ph}{\textrm{\thinspace ph}}
\newcommand{\s}{\textrm{\thinspace s}}

\newcommand{\rg}{\hbox{$\mathrm{\thinspace r_g}$}}

\newcommand{\ctsps}{\hbox{$\cts\s^{-1}$}}
\newcommand{\ergcmps}{\hbox{$\erg\cm\s^{-1}\,$}}
\newcommand{\ergpcmsqps}{\hbox{$\erg\cm^{-2}\s^{-1}\,$}}
\newcommand{\ergps}{\hbox{$\erg\s^{-1}\,$}}
\newcommand{\kmps}{\hbox{$\km\s^{-1}\,$}}
\newcommand{\kmpspMpc}{\hbox{$\kmps\Mpc^{-1}$}}
\newcommand{\pcmsq}{\hbox{$\cm^{-2}\,$}}

\newcommand{\phpcmsqps}{\hbox{$\ph\cm^{-2}\s^{-1}\,$}}

\newcommand{\AFe}{\hbox{$\thinspace A_\mathrm{Fe}$}}

\newcommand{\fwb}{0.37\textwidth}

\begin{document}
\slugcomment{Accepted for Publication in ApJ}   

  \title{\textit{NuSTAR} and \textit{Suzaku} X-ray Spectroscopy of NGC 4151: Evidence for Reflection from the Inner Accretion Disk} 
  \shorttitle{\textit{NuSTAR}/\textit{Suzaku} X-ray Spectroscopy of NGC 4151} 
  \shortauthors{Keck et al.}
  \author{
  M. L. Keck\altaffilmark{1},
  L. W. Brenneman\altaffilmark{2},
  D. R. Ballantyne\altaffilmark{3},
  F. Bauer\altaffilmark{4,5,6},
  S. E. Boggs\altaffilmark{7},
  F. E. Christensen\altaffilmark{8},
  W. W. Craig\altaffilmark{7},
  T. Dauser\altaffilmark{9},
  M. Elvis\altaffilmark{2},
  A. C. Fabian\altaffilmark{10}, 
  F. Fuerst\altaffilmark{11},
  J. Garc\'{\i}a\altaffilmark{2},
  B. W. Grefenstette\altaffilmark{11},
  C. J. Hailey\altaffilmark{12},
  F. A. Harrison\altaffilmark{11},
  G. Madejski\altaffilmark{13},
  A. Marinucci\altaffilmark{14},
  G. Matt\altaffilmark{14},
  C. S. Reynolds\altaffilmark{15},
  D. Stern\altaffilmark{16},
  D. J. Walton\altaffilmark{11,16},
  A. Zoghbi\altaffilmark{17}  
   }	
  \altaffiltext{1}{Institute for Astrophysical Research, Boston University, 725 Commonwealth Avenue, Boston, MA 02215, USA;  \email{keckm@bu.edu}}
  \altaffiltext{2}{Harvard-Smithsonian Center for Astrophysics, 60 Garden Street, Cambridge, MA 02138, USA}
  \altaffiltext{3}{Center for Relativistic Astrophysics, School of Physics, Georgia Institute of Technology, Atlanta, GA 30332, USA}
  \altaffiltext{4}{Instituto de Astrof\'{\i}sica, Facultad de F\'{i}sica, Pontificia Universidad Cat\'{o}lica de Chile, 306, Santiago 22, Chile} 
\altaffiltext{5}{Millennium Institute of Astrophysics, Vicu\~{n}a Mackenna 4860, 7820436 Macul, Santiago, Chile} 
\altaffiltext{6}{Space Science Institute, 4750 Walnut Street, Suite 205, Boulder, Colorado 80301, USA}
\altaffiltext{7}{Space Sciences Laboratory, University of California, Berkeley, CA 94720, USA}
\altaffiltext{8}{DTU Space, National Space Institute, Technical University of Denmark, Elektrovej 327, DK-2800 Lyngby, Denmark}
 \altaffiltext{9}{Dr. Karl Remeis-Observatory and Erlangen Centre for Astroparticle Physics, Sternwartstr. 7, D-96049 Bamberg, Germany}
 \altaffiltext{10}{Institute of Astronomy, University of Cambridge, Madingley Road, Cambridge CB3 0HA, UK}
 \altaffiltext{11}{Space Radiation Laboratory, California Institute of Technology, Pasadena, CA 91125, USA}
 \altaffiltext{12}{Columbia Astrophysics Laboratory, Columbia University, New York, NY 10027, USA}
 \altaffiltext{13}{Kavli Institute for Particle Astrophysics and Cosmology, SLAC National Accelerator Laboratory, Menlo Park, CA 94025, USA}
 \altaffiltext{14}{Dipartimento di Matematica e Fisica, Universit\`{a} degli Studi Roma Tre, via della Vasca Navale 84, I-00146 Roma, Italy}
 \altaffiltext{15}{Department of Astronomy, University of Maryland, College Park, MD 20742-2421, USA}
 \altaffiltext{16}{Jet Propulsion Laboratory, California Institute of Technology, Pasadena, CA 91109, USA}
  \altaffiltext{17}{Department of Astronomy, University of Michigan, 500 Church Street, Ann Arbor, MI 48109-1042, USA}

\begin{abstract}   

We present X-ray timing and spectral analyses of simultaneous 150 $\ks$ \textit{Nuclear Spectroscopic Telescope Array (NuSTAR)} and \textit{Suzaku} X-ray observations of the Seyfert 1.5 galaxy NGC 4151. We disentangle the continuum emission, absorption, and reflection properties of the active galactic nucleus (AGN) by applying inner accretion disk reflection and absorption-dominated models. With a time-averaged spectral analysis, we find strong evidence for relativistic reflection from the inner accretion disk. We find that relativistic emission arises from a highly ionized inner accretion disk with a steep emissivity profile, which suggests an intense, compact illuminating source. We find a preliminary, near-maximal black hole spin $a > 0.9$ accounting for statistical and systematic modeling errors. We find a relatively moderate reflection fraction with respect to predictions for the lamp post geometry, in which the illuminating corona is modeled as a point source. Through a time-resolved spectral analysis, we find that modest coronal and inner disk reflection flux variation drives the spectral variability during the observations. We discuss various physical scenarios for the inner disk reflection model, and we find that a compact corona is consistent with the observed features.   
\end{abstract}

\keywords{accretion, accretion disks $-$ black hole physics $-$ galaxies: active $-$ galaxies: individual (NGC 4151) $-$ galaxies: Seyfert  $-$ X-rays: galaxies}   

\section{Introduction}\label{sec:i}

The dominant features of the X-ray spectra of Seyfert galaxies are a Comptonized continuum, reflection from the accretion disk as well as distant material, and line of sight absorption. Uniquely disentangling these spectral features has been challenging with most X-ray observatories, which primarily have high sensitivity in the $E\sim 0.1-10 \keV$ band. High sensitivity, broadband ($0.5-79 \keV$) observations of Seyfert galaxies enabled by the \textit{Nuclear Spectroscopic Telescope Array} (\textit{NuSTAR}) \citep{Harrison:2013aa} combined with observatories with high sensitivity and energy resolution at lower energies, including \textit{Suzaku} \citep{Mitsuda:2007aa}, allow the continuum, reflection, and absorption features to be definitively separated in Seyfert galaxy X-ray spectra. Deconvolving these features can allow for fundamental properties of active galactic nucleus (AGN) coronae, supermassive black holes (SMBHs), and AGN absorption structures to be determined (e.g., NGC 1365, \citealt{Risaliti:2013aa}, \citealt{Walton:2014aa}; SWIFT J2127.4+5654, \citealt{Marinucci:2014aa}).

The dominant X-ray continuum emission in Seyfert galaxies is commonly thought to be produced by Comptonization of black hole accretion disk thermal emission in a coronal electron plasma (e.g. \citealt{Haardt:1994aa},  \citealt{Reynolds:2003aa}). The corona could be produced by an accretion disk atmosphere or the base of a jet (e.g., \citealt{Markoff:2005aa}). Seyfert galaxies can produce weak jets even though they are radio quiet \citep{Ghisellini:2004aa}. For some AGNs, the corona is compact with a characteristic distance from the accretion disk $D\lesssim10 \rg$ (where $\rg\equiv \rm{GM/c^2}$) as inferred from X-ray micro-lensing, reverberation, and eclipse measurements (e.g., \citealt{Chartas:2009aa}; \citealt{Zoghbi:2012aa}; \citealt{Reis:2013aa}; \citealt{Sanfrutos:2013aa} and references therein).  Because of the compact sizes inferred for AGN coronae,  their detailed geometry has been difficult to determine.

Coronal emission is reprocessed in the black hole accretion disk, which primarily produces relativistically skewed fluorescent Fe K$\alpha$ line emission (rest-frame energy $E=6.4 \keV$ for neutral Fe) and a Compton scattering `hump' at energies $E\gtrsim10 \keV$ (e.g., \citealt{Fabian:1989aa}, \citealt{George:1991aa}). Through measuring the relativistic distortion of the inner accretion disk reflection, important properties of the inner accretion disk and black hole can be determined including the black hole spin (\citealt{Brenneman:2013aa}, \citealt{Reynolds:2013ac}, and references therein).  SMBH spin (which has dimensionless form $a \equiv cJ/GM^2$ where $J$ and $M$ are the black hole angular momentum and mass, respectively) encodes the accretion and merger history of a SMBH \citep{Berti:2008aa}. Building upon the census of the $\sim 22$ SMBH spin measurements currently in the literature (e.g., \citealt{Brenneman:2013aa}, \citealt{Reynolds:2013ac},  \citealt{Walton:2013aa}, and references therein) is crucial for understanding SMBH evolution. SMBH spin also likely plays a critical role in driving AGN outflows and jets that deposit matter and energy in their environment  (e.g., \citealt{Fabian:2012aa}).

Nuclear emission is absorbed in several regions commonly associated with ionized outflows (e.g., \citealt{Elvis:2000aa}),  the broad-line region (\citealt{Antonucci:1993aa}; \citealt{Urry:1995aa}), and/or a molecular torus (e.g, \citealt{Krolik:1988aa}). Absorption and relativistic reflection can produce similar spectral features in the Fe K band. This has led some authors to suggest that the relativistic reflection features actually arise purely from absorption and Compton scattering in material relatively far from the black hole where relativistic effects are negligible (e.g., \citealt{Miller:2009aa}). 

The Seyfert 1.5 galaxy NGC 4151 (z=0.00332, \citealt{de-Vaucouleurs:1991aa}), which is sometimes considered to be the archetypical Seyfert 1 galaxy, is a source in which many AGN phenomena were first characterized (for a review, see \citealt{Ulrich:2000aa}). The source has bolometric luminosity $L_{bol}\sim5\times10^{43} \ergps$ \citep{Woo:2002aa}, SMBH mass $M=4.57_{-0.47}^{+0.57} \times 10^{7} \Msun$ from optical and UV reverberation \citep{Bentz:2006aa}, and corresponding Eddington ratio $L_{bol}/L_{Edd}\sim0.01$. Despite being one of the brightest and most studied AGN, few constraints on the coronal geometry and the black hole spin have been found for the source.

 NGC 4151 has a harder X-ray continuum relative to the average Seyfert AGN. If approximated with a power-law, it has a photon index historically measured to be in the range $\Gamma \sim1.3-1.9$ (e.g., \citealt{Ives:1976aa}; \citealt{Perola:1986aa}; \citealt{Yaqoob:1991aa}; \citealt{Zdziarski:1996aa}; \citealt{Piro:1999aa}; \citealt{Beckmann:2005aa}). 	NGC 4151 was also the first AGN in which a high-energy cutoff was clearly detected (\citealt{Jourdain:1992aa}, \textit{SIGMA}; \citealt{Maisack:1993aa}, \textit{CGRO} OSSE). From an analysis of all \emph{INTEGRAL} data of the source and data from other X-ray observatories during the period 2003 January to 2009 June, \citet{Lubinski:2010aa} found that the coronal emission had an approximately constant X-ray spectral index and Compton $y$ parameter and inferred that the corona had an approximately constant geometry. The X-ray continuum conceivably could arise in the base of a jet, as NGC 4151 has displayed a non-relativistic jet as observed using radio interferometry (\citealt{Wilson:1982aa}; \citealt{Mundell:2003aa}; \citealt{Ulvestad:2005aa}) and characterized with optical (e.g., \citealt{Storchi-Bergmann:2009aa}) and X-ray observations (e.g., \citealt{Wang:2011ac},  \textit{Chandra} ACIS).  

NGC 4151 has shown inconsistent evidence for reflection from the inner accretion disk from X-ray spectroscopy. Spectral analyses of observations taken with instruments with a $E\sim 0.1-10 \keV$ bandpass and CCD resolution have indicated the presence of relativistic reflection in NGC 4151 (e.g.,  \citealt{Yaqoob:1995aa}, \textit{ASCA}; \citealt{Nandra:2007aa}, \textit{XMM-Newton}). However, other analyses find no such evidence for it (e.g., \citealt{Schurch:2003aa}, \textit{XMM-Newton}; \citealt{Patrick:2012aa}, \textit{Suzaku}+\textit{Swift}). The lack of simultaneous broadband ($E\sim1-100 \keV$), high-sensitivity X-ray observations of NGC 4151 combined with the complex absorption (total $N_H \sim 10^{23} \pcmsq$) and soft emission (e.g., \citealt{Armentrout:2007aa}, \citealt{Wang:2011ab}) in the source have also made the unambiguous detection of relativistic reflection challenging.

Recent X-ray reverberation studies of eight archival \textit{XMM-Newton} observations have shown that NGC 4151 has a compact corona surrounding a near-maximally rotating black hole. The discovery of Fe K$\alpha$ reverberation in NGC 4151 by \citet{Zoghbi:2012aa} indicates the presence of broad Fe K$\alpha$ line emission that responds to coronal emission originating from a height above the accretion disk on the order of a few $\rg$. A more recent Fe K$\alpha$ reverberation analysis by \citet{Cackett:2014aa} shows evidence supporting a maximally spinning ($a=0.998$) black hole compared to a low spin black hole ($a=0.1$) and an illuminating source characterized by the lamp post geometry (in which the corona is modeled as a point source on the spin axis; \citealt{Martocchia:2002aa}, \citealt{Miniutti:2003aa}) with height $h = 7.0_{-2.6}^{+2.9}$ $\rg$ ($1\sigma$ errors are given).  

On larger scales, NGC 4151 exhibits partial-covering, neutral material (e.g., \citealt{Holt:1980aa}) and ionized outflows (e.g., \citealt{Wang:2011ab} and references therein) viewed in absorption.  The source has displayed absorption column density variability on time-scales as short as $\Delta t\sim2$ days that indicates that the absorbing material is located in the broad-line region (\citealt{Puccetti:2007aa}, \citealt{de-Rosa:2007aa}).  The absorption has also been characterized as having an ionized\footnote{Ionization is parameterized with the ionization parameter, which is defined as the ratio of the ionizing X-ray flux to the gas density: $\xi \equiv \frac{4\pi F_x}{n}$.} ($\xi \sim 500 \ergcmps$), outflowing ($v_{out}\sim 500 \kmps$)  component suggestive of an accretion disk wind (\textit{Chandra} HETG; \citealt{Kraemer:2005aa}) as well as a highly ionized ($\xi \sim 10^4-10^5 \ergcmps$) and blue-shifted ($v_{out} \sim 0.1c$) `ultra-fast outflow' (UFO) component that might be associated with an accretion disk wind or a weak jet \citep{Tombesi:2011aa}. 

In this paper, we present timing and spectral analyses of simultaneous 150 $\ks$ \textit {NuSTAR} and \textit{Suzaku} X-ray observations of NGC 4151 in a moderately bright state focused on deconvolving the continuum, reflection, and absorption features. In a separate paper, we will present a reverberation analysis of these observations (Zoghbi et al., in prep.). This paper is structured as follows: \S\ref{sec:odr} summarizes the observations and data reduction; \S\ref{sec:ta} discusses the timing analysis; \S\ref{sec:sa} presents time-averaged (\S\ref{sec:taa}) and time-resolved (\S\ref{sec:tra}) spectral analyses; \S\ref{sec:d} discusses systematic errors and the physical implications of our results; and \S\ref{sec:c} presents our conclusions. 

\section{Observations and Data Reduction} \label{sec:odr}

\begin{figure}
 \centering
\includegraphics[width=0.35\textwidth, angle=270]{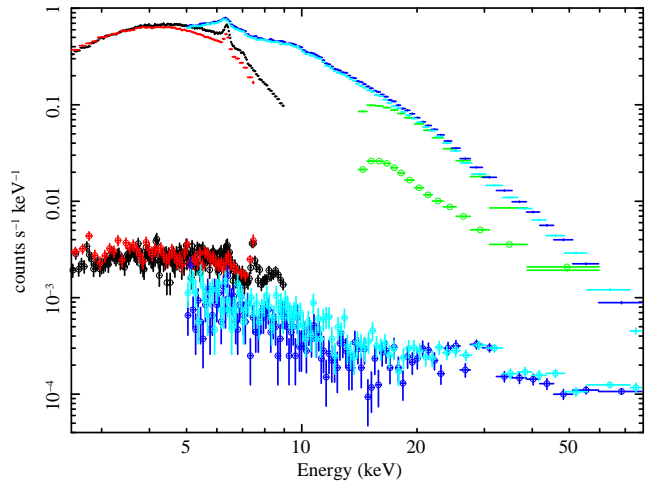}
\caption{Source and background spectra for the XIS-FI (shown in black), XIS-1 (red), PIN (green), FPMA (blue), and FPMB (light blue) instruments. Source count rates are marked with error bars, and background count rates are marked with error bars in circles. For all instruments, source count rates exceed background count rates except for the PIN instrument above $E\sim40 \keV$. (A color version of this figure is available in the online journal.)
  \label{fig:spectra}}
\end{figure}

\textit{Suzaku} and \textit{NuSTAR} observed NGC 4151 from November 11-14, 2012.  \textit{Suzaku} observations were taken with NGC 4151 placed at the nominal X-ray Imaging Spectrometer (XIS) pointing position. The \textit{Suzaku} observation has observation ID 707024010, and the \textit{NuSTAR} observation is composed of three pointings with observation IDs 60001111002, 60001111003, and 60001111005. \textit{Suzaku} and  \textit{NuSTAR} observation information is given in Table~\ref{tab:obsinfa}. Source and background spectra provide a source-dominated view of the $2.5-79 \keV$ emission from NGC 4151 as shown in Figure~\ref{fig:spectra}. We reprocessed all data using \texttt{HEAsoft} version 6.16, which includes {\tt XSPEC} v12.8.2 \citep{Arnaud:1996aa} and \texttt{xronos} v5.22.

\begin{deluxetable}{ccccccc}
   \tablecaption{\textit{Suzaku} and \textit{NuSTAR} observation information \label{tab:obsinfa}}
   \tablehead{ \colhead{Instrument} & \colhead{Band\tablenotemark{a}} &   \colhead{Time}& \colhead{Count Rate} & \colhead{Counts}  \\
    \colhead{ } & \colhead{(\keV)} &   \colhead{(\ks)}& \colhead{(\ctsps)} & \colhead{} }
   \startdata
     \textit{Suzaku}   XIS-0+3  & $2.5-9$    & 150   & $6.180 \pm 0.007$ & 929099\\
     \textit{Suzaku}   XIS-1   &  $2.5-7.5$ &   150   & $2.444 \pm 0.004$ & 367408 \\
      \textit{Suzaku}     PIN  & $14-60$    & 141   & $0.985 \pm 0.003$ & 138881 \\
    \textit{NuSTAR}  FPMA\tablenotemark{b}  &  $5-79$    & 141   & $5.154 \pm 0.006$  & 723597 \\
    \textit{NuSTAR}    FPMB\tablenotemark{b}  & $5-79$  & 141   & $4.915 \pm 0.006$  & 690323 \\
\enddata
\tablecomments{Counts and count rates are background subtracted.}
    \tablenotetext{a}{Column gives energy bands used in the spectral analysis.}
    \tablenotetext{b}{\textit{NuSTAR} data corresponds to the co-added \textit{NuSTAR} observations.}
\end{deluxetable}%

\subsection{Suzaku}

We produced \textit{Suzaku} data from the XIS \citep{Koyama:2007aa} (XIS 0, XIS 1, and XIS 3) using the \texttt{aepipeline} script following the \textit{Suzaku} ABC Guide\footnote{ \label{notea} \url{http://heasarc.gsfc.nasa.gov/docs/suzaku/analysis/abc/}} with version 22 of the \textit{Suzaku} software and the XIS \texttt{CALDB} release of 2015 January 5, which includes updated XIS gain files. We extracted  \textit{Suzaku} XIS data products (i.e.  spectra,  response files, and light curves) using \texttt{xselect} and \texttt{xisresp} from circular, $170 \arcsec$ radius regions centered on the source. For each observation, background spectra were extracted from as much of the chip as possible. Source and background regions excluded all contaminating sources in the field of view (primarily the BL Lac object 1E1207.9+3945 and the LINER NGC 4156) and calibration sources in the corners of the chip. 

 We rebinned \textit{Suzaku} XIS spectra and response files from 4096 to 2048 channels to expedite spectral fitting without compromising the energy resolution of the data. We then co-added data from the XIS front-illuminated (XIS-FI) instruments, XIS 0 and XIS 3, using the \texttt{addascaspec} script to maximize signal-to-noise (S/N). We rebinned XIS source spectra into channels with a minimum of 25 counts per bin using \texttt{grppha} to allow for robust $\chi^2$ fitting.  In the spectral analysis, we apply a cross-normalization constant to all detectors relative to the co-added XIS-FI data. XIS-1 fits with a cross-normalization of $0.96\pm0.01$, which is lower than the nominal XIS-1/XIS-FI cross-normalization\footnote{\label{noteb} \url{http://heasarc.gsfc.nasa.gov/docs/suzaku/analysis/watchout.html}}, $1.019 \pm 0.010$. However, our value is reasonable given that the cross-normalization depends on the exact spectral shape of the source and exactly where the source is on the chip (K. Mukai, private communication). 

Because \textit{Suzaku} and \textit{NuSTAR} observed NGC 4151 in a relatively bright state, we assessed the influence of pileup in the XIS data (note that \textit{NuSTAR} does not suffer from pileup) using the \texttt{pileest} script. We found that, for XIS 0, 1, and 3,  the central $\sim 30 \arcsec$ has a pileup fraction $3 \%\lesssim f_{pl} \lesssim 10 \%$ assuming a grade migration parameter of 0.5. We found that this minor pileup fraction does not influence our conclusions in the timing and spectral analyses on the spectral properties of NGC 4151 as described in Appendix~\ref{sec:appA}.

 We reprocessed the \textit{Suzaku} Hard X-ray Detector (HXD) PIN instrument  (\citealt{Takahashi:2007aa})  data using the {\tt aepipeline} script with the 2011 September 15 HXD calibration files. We reduced and extracted it following the \textit{Suzaku} ABC Guide\footnotemark[19]. We estimated the PIN non-X-ray background using the version 2.2 `tuned' non-X-ray background (NXB) event file for November 2012 as well as response and flat-field files for epoch 11 data. We estimated the PIN cosmic X-ray background with {\tt XSPEC} simulations using the model of \citet{Boldt:1987aa}.  We rebinned PIN data to a S/N of 5 using \texttt{ISIS} \citep{Houck:2000aa} to facilitate $\chi^2$ fitting. Three percent systematic errors were added to PIN data as recommended by the \textit{Suzaku} team to account for uncertainties present in the NXB data. We allow the cross-normalization factor of PIN relative to the XIS-FI instruments to vary in the spectral analysis to improve upon the agreement between the PIN and FPMA/FPMB spectra found with the nominal value\footnotemark[20], $1.164\pm0.014$. It fits to $1.26\pm0.01$. 

\subsection{NuSTAR}

We reprocessed \textit{NuSTAR} focal plane module A and B (FPMA and FPMB) data using the \textit{NuSTAR} Data Analysis Software (\texttt{NuSTARDAS}) version 1.4.1, using the \textit{NuSTAR} {\tt CALDB} from 2014 July 1.  \texttt{NuSTARDAS} performs standard data calibration and data screening, including the elimination of data collected when the Earth occulted NGC 4151 and when \textit{NuSTAR} was in the South Atlantic Anomaly. For each of the \textit{NuSTAR} observations, we extracted source data products from a $70\arcsec$ radius circular region centered on NGC 4151, and we extracted background spectra from $70 \arcsec$ radius circular regions on the same detector chosen to avoid contamination and detector edges. Background regions were separated by at least $4\arcmin$ from NGC 4151 and $\gtrsim7\arcmin$ from 1E1207.9+3945 and NGC 4156, which exclude $\gtrsim98\%$ and $\gtrsim99\%$ of the enclosed counts for a point source, respectively \citep{Harrison:2013aa}. Spectra from the three \textit{NuSTAR} observations were co-added and analyzed separately for FPMA and FPMB.  We rebinned FPMA and FPMB spectra into channels with a minimum S/N of 5 to allow for $\chi^2$ fitting. We allow the FPMA and FPMB cross-normalization constants relative to XIS-FI to fit freely in the spectral analysis, and they fit to $1.04\pm0.01$ and  $1.08\pm0.01$, respectively. 

\section{Timing Analysis} \label{sec:ta}

\begin{figure*}
 \centering
\includegraphics[width=0.6\textwidth,  angle=270]{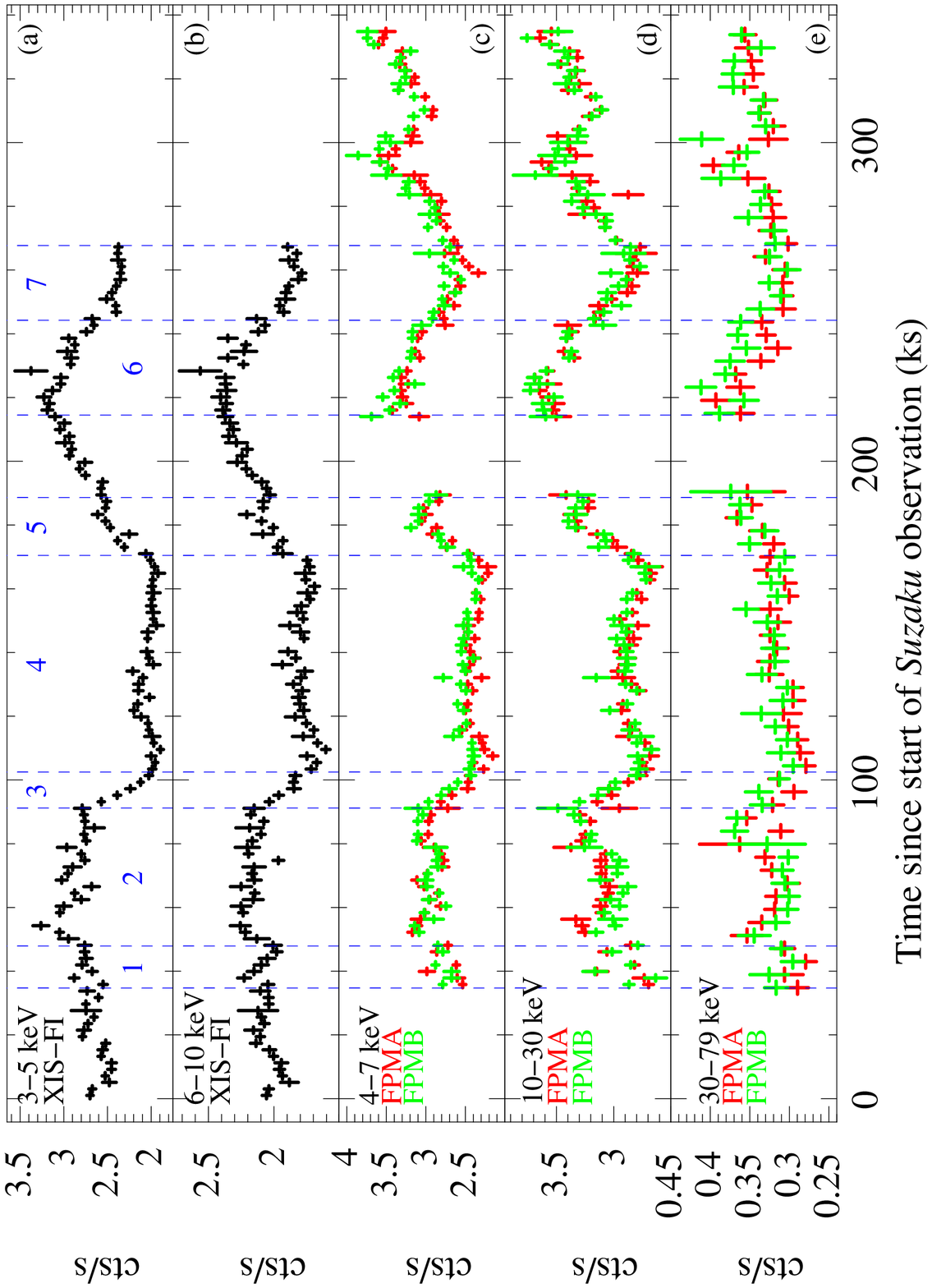}
\includegraphics[width=0.6\textwidth,  angle=270]{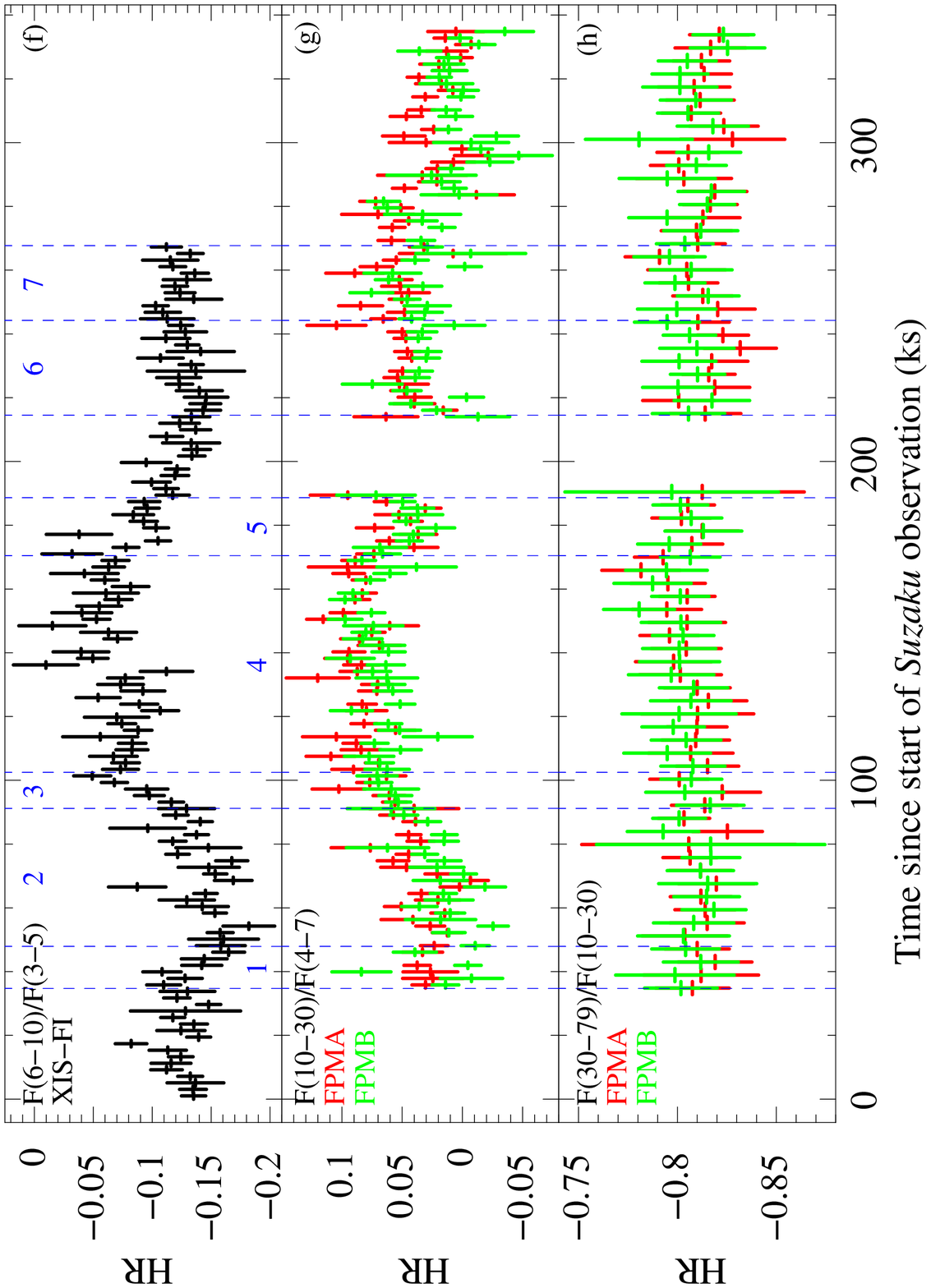}
 \caption{Light curves and hardness ratios with $2048 \s$ bins (except for panels (e) and (h), which are shown with $4096 \s$ bins to better show the variability of the $30-79 \keV$ flux). Note the reduction in the broad-band flux coincident with a general increase in spectral hardness seen during a $\Delta t\sim 15\ks$ period labelled interval 3. Panels (a) and (b) show the \textit{Suzaku} XIS-FI $3-5 \keV$ and $6-10 \keV$ light curves as labelled. Panels (c), (d), and (e) show \textit{NuSTAR} light curves in the $4-7 \keV$, $10-30 \keV$, and $30-79 \keV$ bands, respectively. Panels (f), (g), and (h) show the hardness ratios  $F(6-10 \keV)/F(3-5 \keV)$, $F(10-30 \keV)/F(4-7 \keV)$, and $F(30-79 \keV)/F(10-30 \keV)$, respectively.  Vertical, blue dashed lines indicate time intervals 1-7 analyzed in the time-resolved analysis as numbered. (A color version of this figure is available in the online journal.)  \label{fig:hrnusu}}
\end{figure*}

 The \textit{Suzaku} XIS and \textit{NuSTAR} light curves and hardness ratios have small but significant variability as shown in Figure ~\ref{fig:hrnusu}.  Notably, a drop in flux coincident with an increase in spectral hardness is seen during a $\Delta t\sim 15\ks$ period labelled interval 3. During interval 3, the $3-5 \keV$ flux drops by a factor of $\sim1.4$ and the $10-30\keV$ flux decreases by a factor of $\sim1.2$. At the same time, the hardness ratio (HR) between the $6-10 \keV$ flux and the $3-5 \keV$ flux increases by $\sim5\sigma$. Between intervals 4 and 6 ($\Delta t\sim 45\ks$), the $3-5 \keV$ flux rises by a factor of $\sim1.6$, the $10-30\keV$ flux increases by a factor of $\sim1.3$, and the $F(6-10 \keV)/F(3-5 \keV)$ HR decreases by $\sim4\sigma$. From the beginning of interval 6 to the end of interval 7 ($\Delta t\sim 40\ks$), the $3-5 \keV$ flux drops by a factor of $\sim1.4$, the $30-79\keV$ flux decreases by a factor of $\sim1.3$, and the $F(6-10 \keV)/F(3-5 \keV)$ HR remains approximately constant. 

\begin{figure}
 \centering
\includegraphics[width=0.35\textwidth, angle=270]{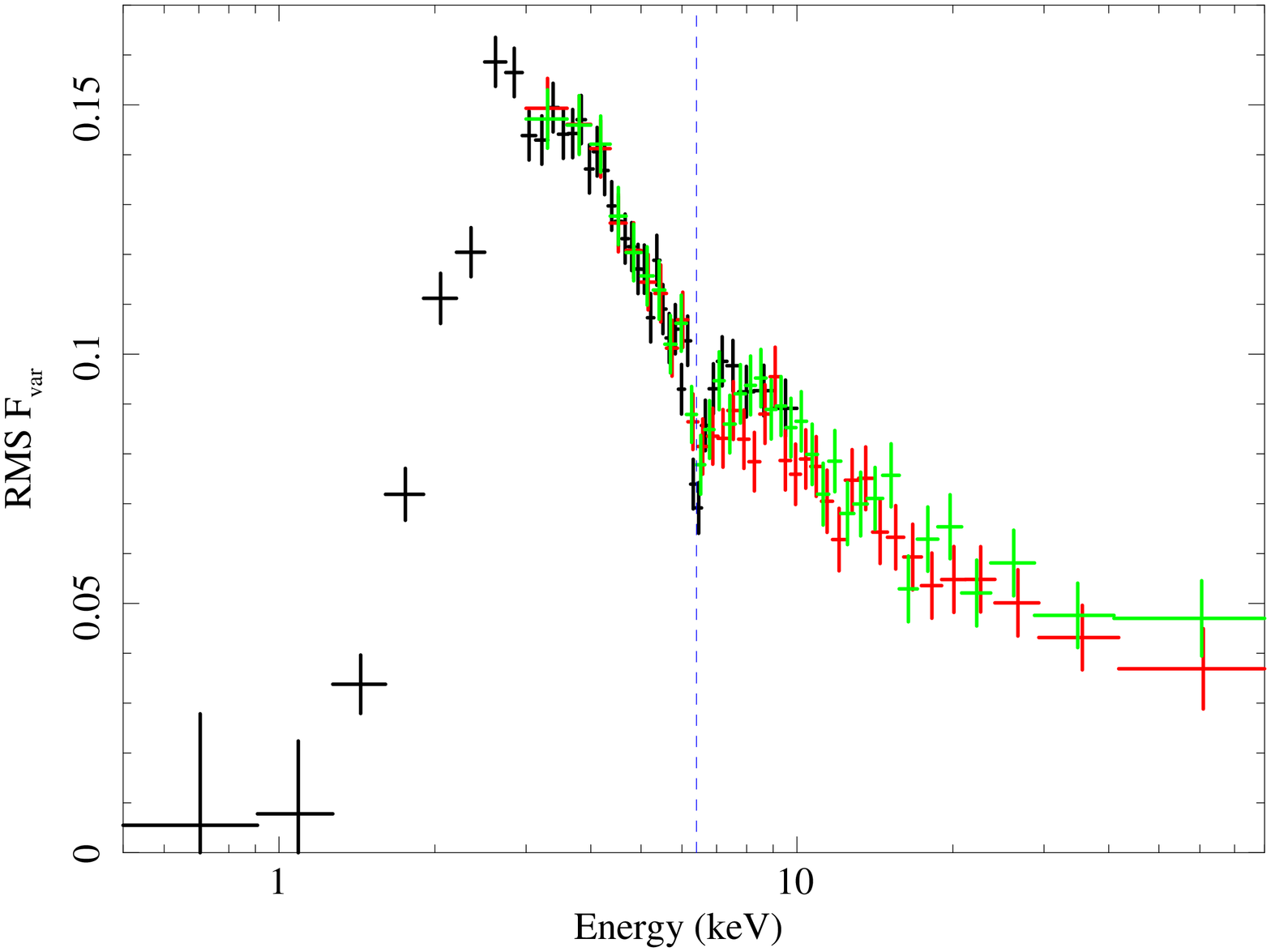}
 \caption{The RMS fractional variability of the data calculated with 4000 $\s$ bins. XIS-FI is in black, FPMA is red, and FMPB is green (XIS-1 is not shown for clarity). Data are included in the energy range $0.7-10 \keV$ for XIS-FI and $3-79 \keV$  for FPMA and FPMB. The dashed blue line indicates the rest-frame energy of the neutral Fe K$\alpha$ emission line ($6.4 \keV$).  (A color version of this figure is available in the online journal.)  \label{fig:rmsfvar}}
\end{figure}

The root-mean-square fractional variability (RMS $F_{var}$) is shown in Figure~\ref{fig:rmsfvar}. The overall source variability drops dramatically  below $E\sim3 \keV$, where circumnuclear and scattered emission dominate the absorbed emission from the AGN (e.g., \citealt{Wang:2011ab}).  There is also an overall drop in variability with increasing energy above $E\sim 3\keV$, which is typical of Seyfert AGN, where the more variable model components tend to dominate at softer energies. The sharp, relatively narrow dip at $6.4 \keV$ is an indication that the neutral, narrow component of the Fe K$\alpha$ line is not varying much relative to the surrounding continuum during the observation. This is expected of the neutral, narrow reflection component because it likely originates relatively far from the nucleus. 

\section{Spectral Analysis} \label{sec:sa}

We perform time-averaged and time-resolved analyses to characterize the broadband X-ray spectral features and variability of NGC 4151.  We first undertake a time-averaged spectral analysis (i.e., jointly fitting the full XIS-FI, XIS-1, PIN, FPMA, and FPMB spectra) to understand the components present in the source spectrum using the high signal-to-noise (S/N) of the full data. We then carry out a time-resolved analysis to characterize spectral variability that is averaged out in the full observations, which may include variations in the coronal emission (which can vary on hour-long time scales in AGN; e.g., \citealt{Markoff:2005aa}), the inner accretion disk reflection emission (which can vary on the order of hours in NGC 4151; e.g., \citealt{Zoghbi:2012aa})  and absorption features (which can vary on the order of days in NGC 4151; e.g., \citealt{Puccetti:2007aa}). 

To carry out the spectral analyses, we use {\tt XSPEC} with a standard cosmology of $H_0 = 70 \kmpspMpc$, $\Omega_M = 0.27$, and $\Omega_\Lambda = 0.73$ \citep{Komatsu:2011aa}. We use X-ray cross-section values from \citet{Verner:1996aa}  and solar abundance values from \citet{Wilms:2000aa}. We include data in the energy ranges indicated in Table~\ref{tab:obsinfa}.  

For the spectral analysis, we apply an inner disk reflection model and an absorption-dominated model. In brief, the inner disk reflection model includes a partial-covering, neutral absorption component (which we refer to as PC1), a distant reflection component (DRC), a cut-off power-law component (PLC), and an inner disk reflection component (IDR). The absorption-dominated model is identical to the inner disk reflection model except that the IDR component is replaced with another partial-covering absorber (PC2). For the time-averaged analysis, we link all parameters between data sets, and we fix redshifts of all model components to the systemic redshift. We model Milky way absorption with {\tt Tbabs} \citep{Wilms:2000aa}, and we fix the Milky Way absorption column density to the weighted-average value for NGC 4151 from the LAB survey \citep{Kalberla:2005aa}, $N_H = 2.3 \times 10^{20} \pcmsq$. We determine $90\%$ confidence intervals for one interesting parameter for all parameter values presented in the time-averaged and time-resolved analyses using the \texttt{XSPEC} error algorithm unless noted otherwise.

\subsection{Time-Averaged Analysis} \label{sec:taa}

\begin{figure*}
 \centering
\includegraphics[width=0.37\textwidth, angle=270]{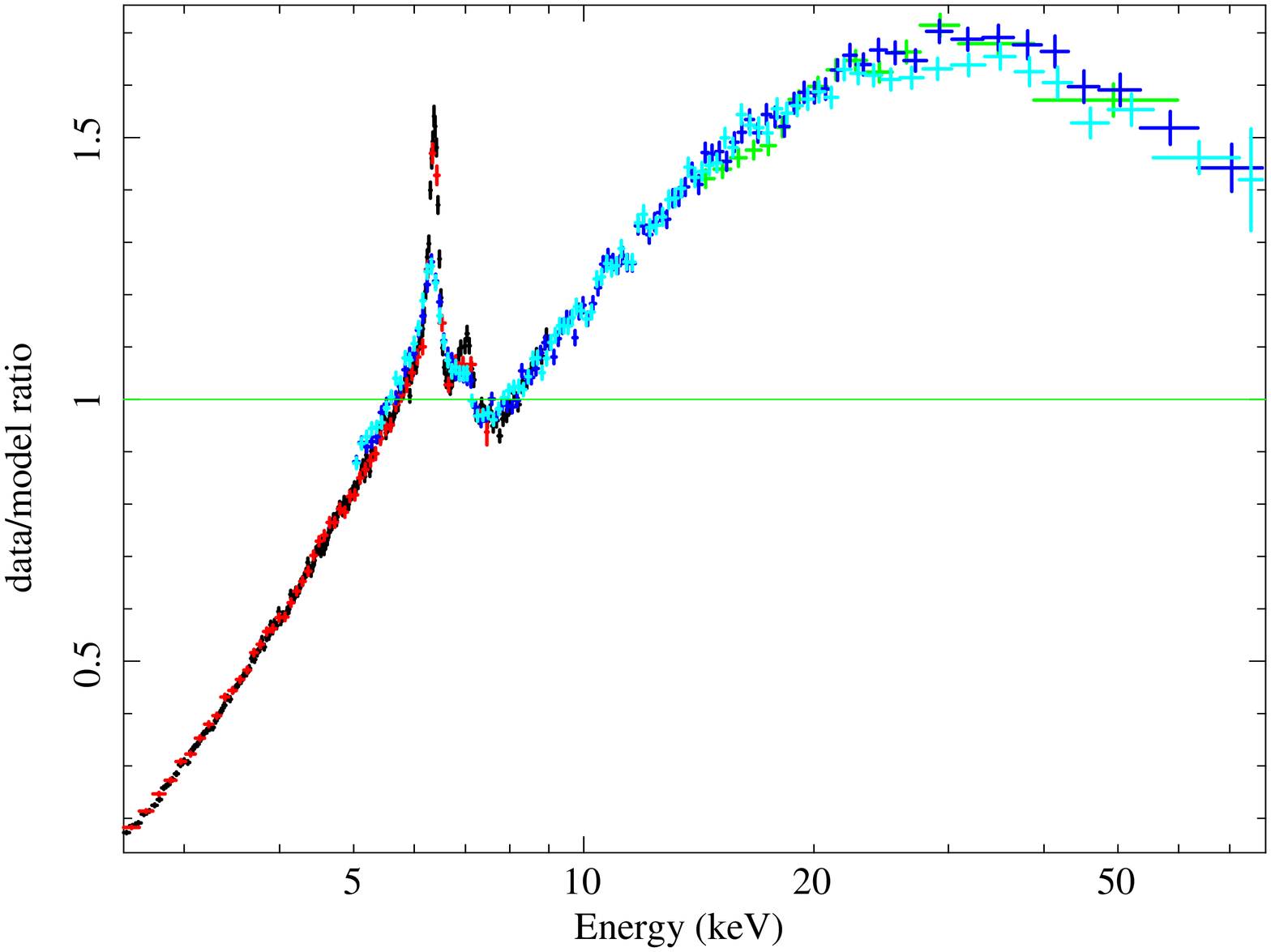}
\includegraphics[width=0.37\textwidth, angle=270]{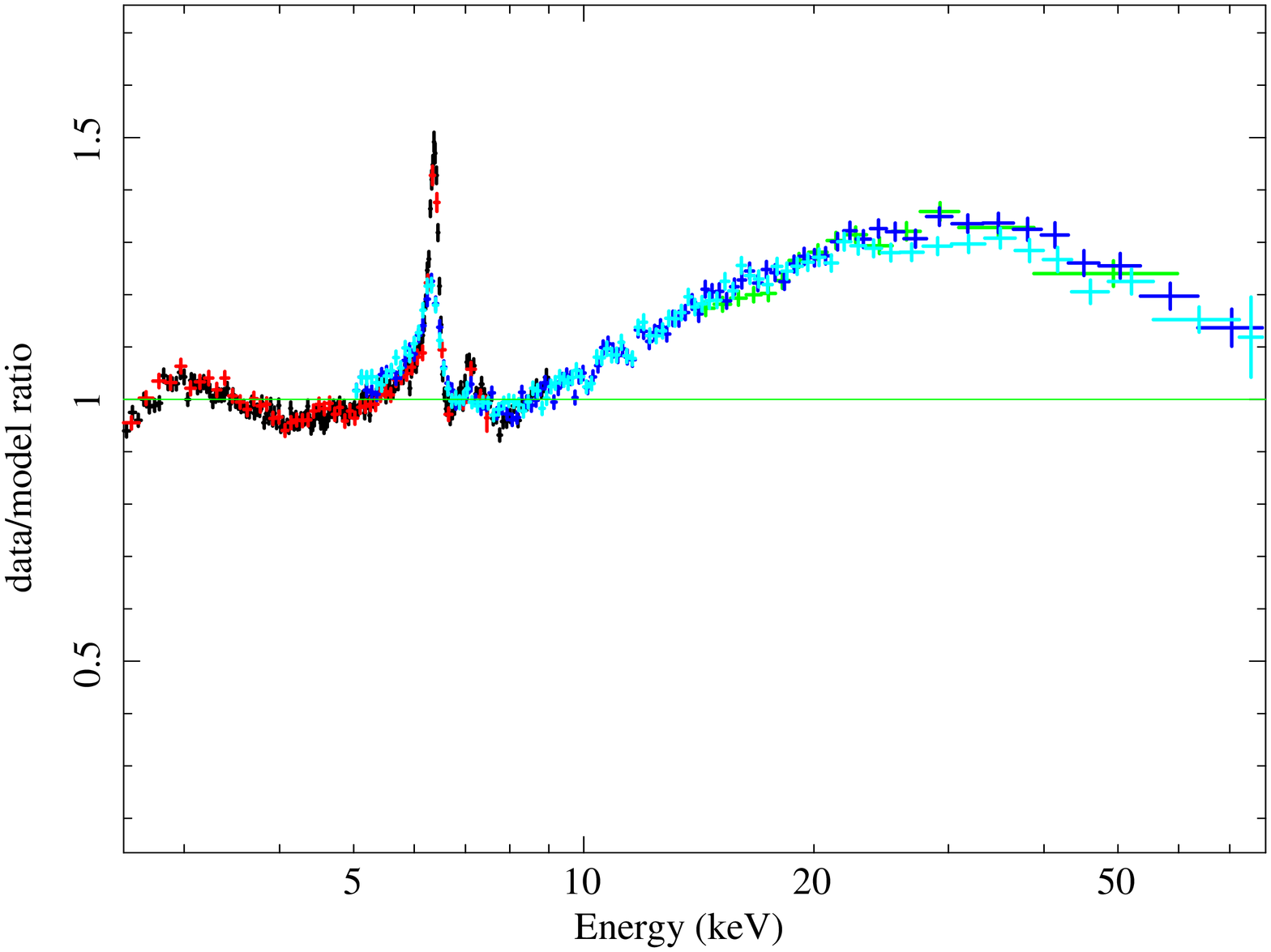}
\includegraphics[width=0.37\textwidth, angle=270]{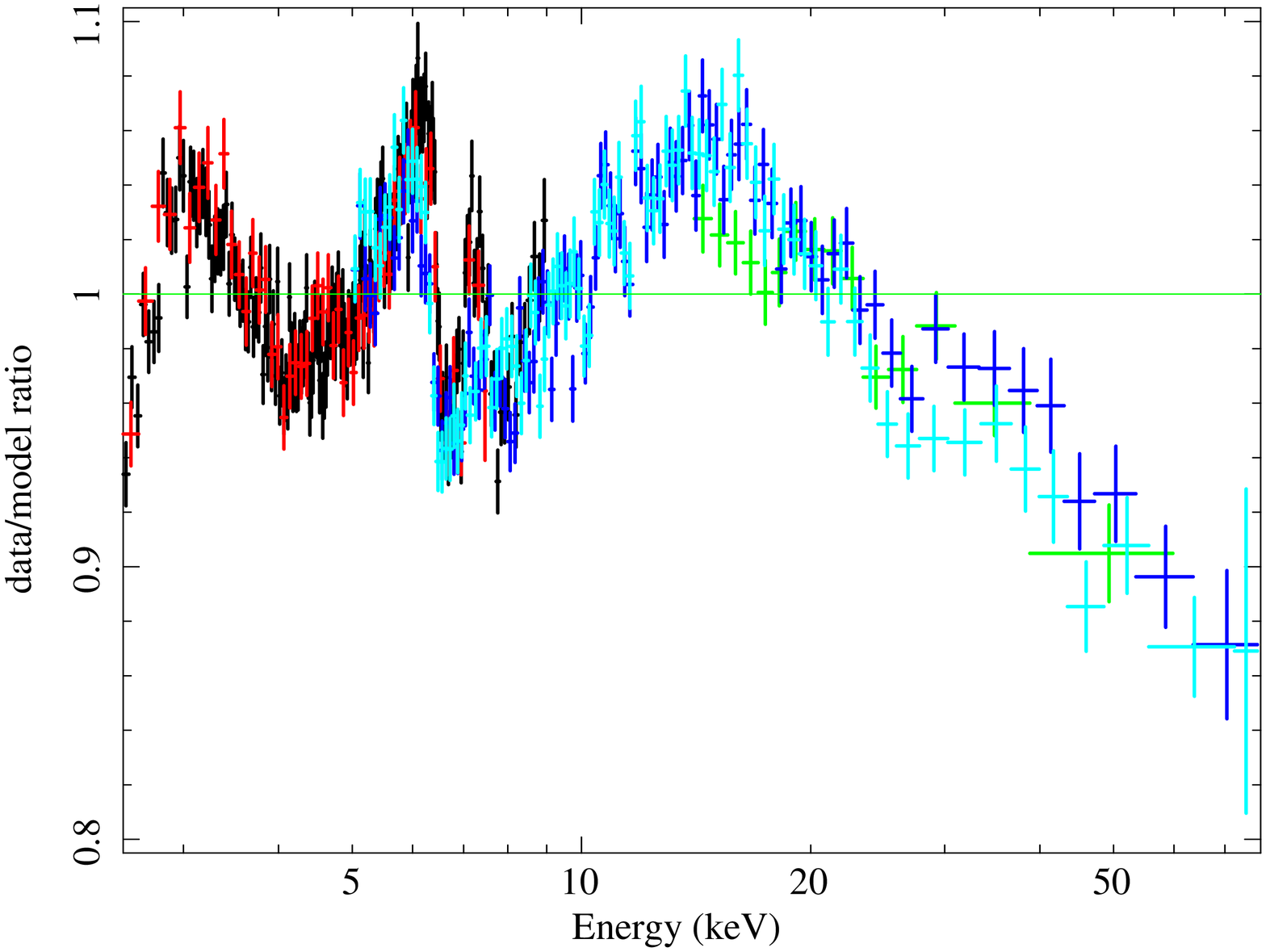}
\includegraphics[width=0.37\textwidth, angle=270]{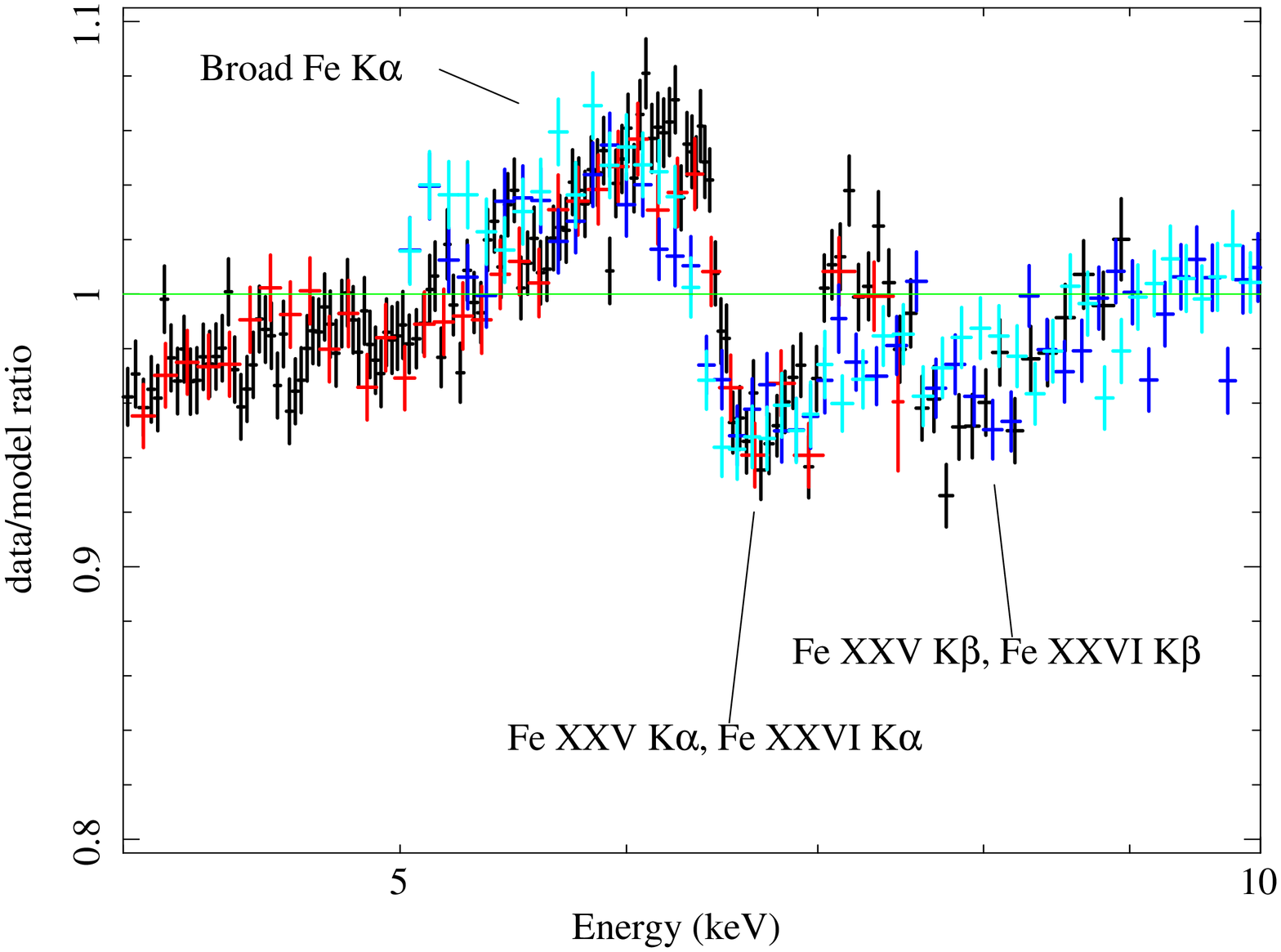}
\includegraphics[width=0.37\textwidth, angle=270]{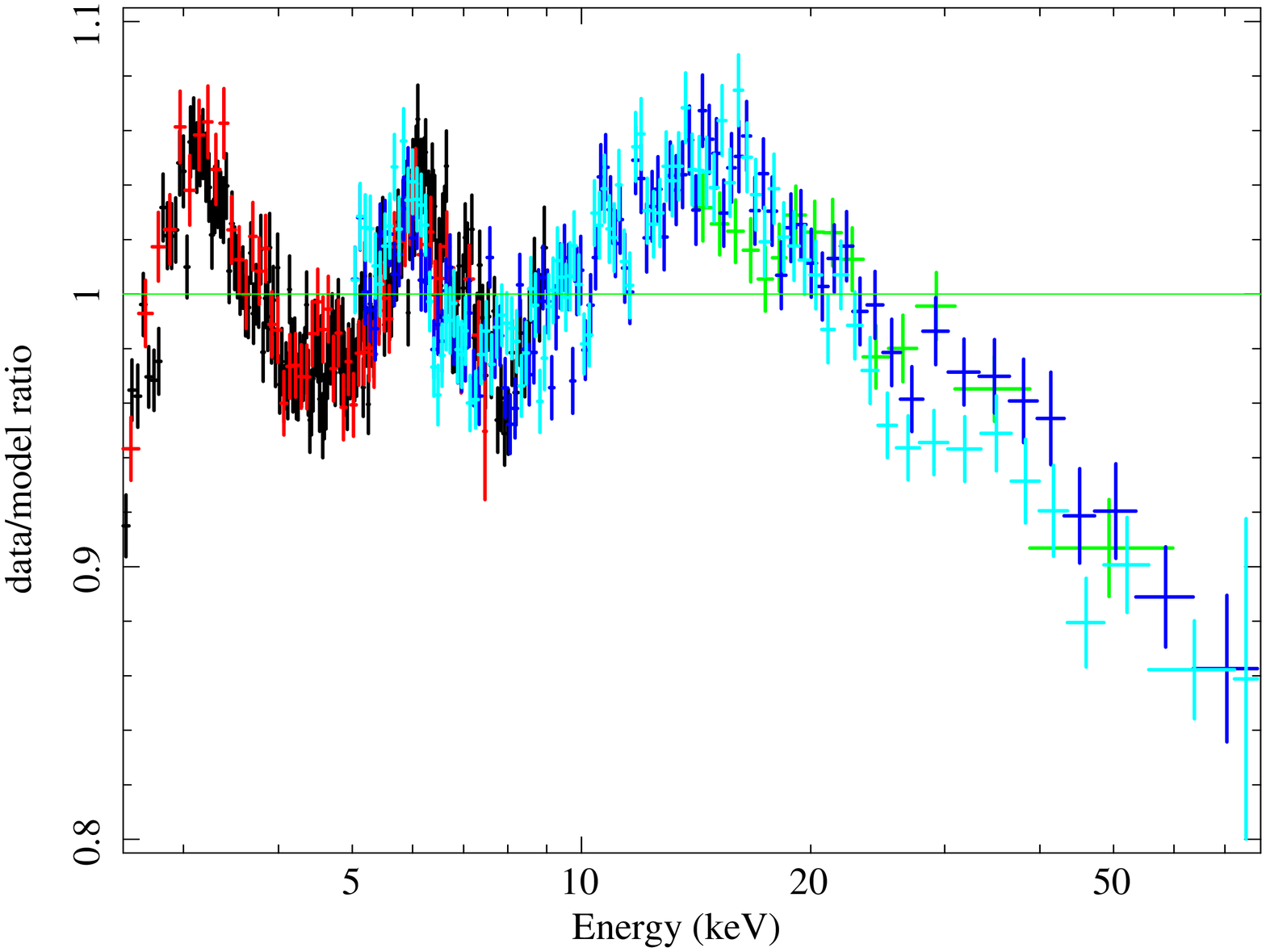}
\includegraphics[width=0.37\textwidth, angle=270]{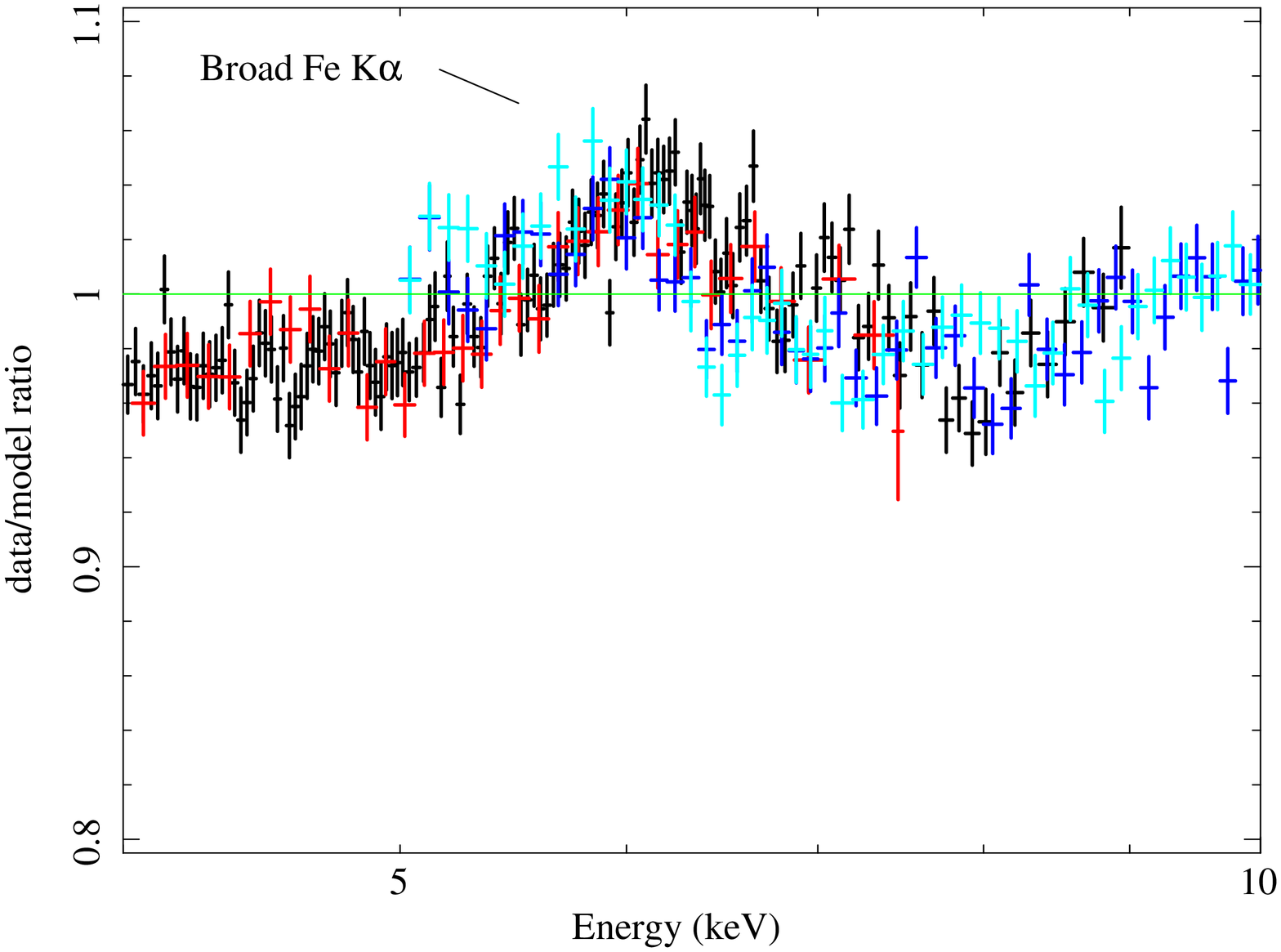}
 \caption{ \textit{Top Left:} Data/model ratios of a power-law ($\Gamma=1.75$) absorbed only by gas in the Milky Way that demonstrate the strong absorption in the source. \textit{Top Right}: Data/model ratios after adding a partial covering, neutral absorber and refitting that show the spectral features of relatively neutral reflection. \textit{Middle Left}: Best-fit ratios after additionally including a neutral reflection component with free Fe abundance and refitting with $\Gamma=1.75$ fixed. \textit{Middle Right}: Same as \textit{Middle Left}, focused on the ratios in the $4-10 \keV$ band. Notice the probable broad Fe K$\alpha$ line as well as the blended \ion{Fe}{25} and \ion{Fe}{26} absorption lines. \textit{Bottom Left}:  Best-fit ratios after additionally including two warm absorber components and refitting with $\Gamma=1.75$ fixed. \textit{Bottom Right}:  Same as \textit{Bottom Left}, focused on the ratios in the $4-10 \keV$ band. For plotting purposes only, spectral bins have either a significance of $80\sigma$ or a maximum of $80$ spectral channels for each panel. Also, XIS-FI data is black, XIS-1 is red, PIN is light green, FPMA is dark blue, and FPMB is light blue. (A color version of this figure is available in the online journal.)  \label{fig:taModMot}}
\end{figure*}

 To develop the time-averaged model, we first characterize the continuum absorption and emission features in the $2.5-4.0 \keV$, $7.5-10 \keV$, and $50-79 \keV$ bands. We model the coronal continuum emission with a cut-off power-law with photon index $\Gamma$ and cut-off fixed at $E_{cut}=1000\keV$.  Data/model ratios in Figure~\ref{fig:taModMot} illustrate the presence of a strong absorption cut-off at energy $E\sim2-3 \keV$ indicative of a relatively high line-of-sight column density ($N_H \sim 10^{23} \pcmsq$) as well as a prominent, narrow Fe K$\alpha$ emission line at $6.4 \keV$ and Compton hump indicative of reflection. 
 
  We apply a partial-covering, neutral absorption component (\texttt{partcov*zTbabs}; \citealt{Wilms:2000aa}).  This absorber greatly improves the fit (see Table~\ref{tab:taModMot}) and satisfactorily accounts for the low-energy continuum ratios as shown in Figure~\ref{fig:taModMot}. We then model the prominent, narrow Fe K$\alpha$ emission line and Compton hump as arising by reflection of cut-off power-law emission from Compton-thick, plane-parallel, and relatively neutral (i.e., $\xi = 1 \ergcmps$)  material with the \texttt{xillver} reflection model \citep{Garcia:2013aa}. We specifically employ \texttt{xillver-Ec3}\footnote{available at \url{http://hea-www.cfa.harvard.edu/~javier/xillver/}}, which allows for a variable cut-off energy in the range $E=20-1000 \keV$. We fix the incident cut-off power-law photon index and cut-off energy of this component self-consistently to those of the cut-off power-law component, and we additionally fix its inclination to $i=60 \deg$. As shown in the bottom panels of Figure~\ref{fig:taModMot}, the added \texttt{xillver} component significantly reduces the ratios towards unity. The remaining residuals indicate the presence of blended Fe XXV and Fe XXVI K$\alpha$ ($E=6.67 \keV$ and $E=6.97 \keV$, respectively) and K$\beta$ ($E=7.88 \keV$ and $E=8.25 \keV$, respectively) absorption lines as well as broad emission in the Fe K band and convex curvature above $E\sim 10 \keV$.
 
  \begin{deluxetable}{cccc}
 \tabletypesize{\footnotesize}
\tablecaption{$\chi^2$ and $\nu$ after addition of each time-averaged model component  \label{tab:taModMot}}
\tablehead{Additional component & \colhead{$\chi^2$} &$\nu$  &$\chi^2/\nu$ }
\startdata
    cut-off power-law & $264344$ & $4120$ & $64.2$ \\   
    partial covering absorber & $15626$ & $4118$ & $3.80$ \\  
    distant reflector\tablenotemark{a} & $6264$ & $4112$ & $1.52$   \\
    warm absorber 1 &  $5581$ & $4110$ & $1.36$  \\
    warm absorber 2 & $5513$ & $4108$ & $1.34$  \\
    inner disk reflector    & $4453$ &$4101$  & $1.09$    
  \enddata
  \tablecomments{Each added model component provides a significant improvement to the fit.} 
  \tablenotetext{a}{Values includes a change $\Delta\chi^2/\Delta\nu =-39/-4$ from allowing the instrument cross-normalizations to fit freely.}
\end{deluxetable}

To account for the unmodeled absorption lines in the Fe band, we additionally apply two separate {\tt XSTAR} absorption table models, which we refer to as warm absorbers one (WA1) and two (WA2). WA1 is parameterized by its column density in the range $N_H=10^{22}-10^{24} \pcmsq$ and ionization parameter $\xi$ in the range $\mathrm{log}(\xi/\ergcmps)= 2.5-4.5$. The absorber has turbulent velocity $v_t = 3000 \kmps$, which is appropriate for a highly ionized absorber \citep{Tombesi:2011aa}.  We fix the Fe abundance to one. WA2 is parameterized by its column density in the range $N_H=10^{20}-10^{24} \pcmsq$ and $\xi$ in the range $\mathrm{log}(\xi/\ergcmps)= 0.0-4.0$. It has a turbulent velocity $v_t = 200 \kmps$. The Fe abundance is hard-wired at one. We limit $\xi$ to be in the range $\mathrm{log}(\xi/\ergcmps)= 2.5-4.0$. For both WA1 and WA2, $\xi$ is sampled with resolution $\Delta \mathrm{log}(\xi/\ergcmps) = 0.2$, which is important for accurately modeling the absorption opacities and, consequently,	 determining the black hole spin \citep{Reynolds:2012ab}. Having accounted for these spectral features, we develop separate inner disk reflection and absorption-dominated models to account for the remaining broad residuals. 

\subsubsection{Inner disk reflection model}\label{sec:taaidr}

We first characterize the broad residuals in the Fe K-band phenomenologically. We begin by applying a broad Gaussian. The line fits with equivalent width $EW=66_{-15}^{+16} \eV$, $\sigma=0.39\pm0.04\keV$, and line energy $E=6.20 \pm 0.05 \keV$. The fit improves by $\Delta\chi^2 / \Delta \nu = -341 / -3$. The measured line energy and width are suggestive of broadened Fe K$\alpha$ emission. To test for an origin from the inner accretion disk, we replace the Gaussian with a model of line emission from the inner accretion disk ({\tt relline}; \citealt{Dauser:2010aa}) with an emissivity profile $\epsilon\propto r^{-q}$ with $q=3$ fixed. The {\tt relline} component fits with  line energy $E=6.52 \pm 0.01 \keV$,  $EW=111_{-22}^{+16} \eV$, inclination $i=6 \pm 1 \deg$, and spin $a>0.99$. The addition of the broad Fe line is statistically motivated by a reduction in $\chi^2$ by $\Delta\chi^2 / \Delta \nu = -354 / -4$.

To account for emission from the inner accretion disk self-consistently (i.e., by including the full reflection spectrum, including the Compton hump, Fe K$\alpha$ and K$\beta$ emission lines, and the Fe K absorption edge), we include an inner disk reflection component (IDR) composed of the {\tt xillver} model convolved with the {\tt relconv} relativistic convolution model  \citep{Dauser:2010aa}. Specifically, we use an optimized version {\tt relconvf} (Jeremy Sanders, private communication). \texttt{relconv} allows the black hole spin, $a$, to fit freely in the range $-0.998 \le a \le 0.998$. We further allow the inner disk reflection component to have a broken power-law emissivity profile that transitions from inner emissivity index $q_1$ to outer emissivity index $q_2$ at a radius $r_{br}$.  {\tt relconv} additionally assumes an inner accretion disk that is geometrically thin, optically thick, and aligned perpendicular to the spin axis.

We assume that the inner disk radius ($r_{in}$) is fixed at the innermost stable circular orbit (ISCO) radius and that the outer disk radius is fixed at $r_{out}=400 \rg$. This value should be unconstrained assuming $q_2 \gtrsim 3$ as expected for rapidly spinning black holes (e.g., \citealt{Wilkins:2011aa}). We assume an isotropic emission distribution as discussed in \citet{Svoboda:2009aa}. We further assume that the AGN is chemically homogeneous as discussed in, e.g.,  \citet{Reynolds:2012ab} and \citet{Walton:2013aa}, so we fix the Fe abundances of the distant reflector and IDR together. We additionally limit the Fe abundance to be in the range $0.5 < \AFe < 5.0$ in order to keep the Fe abundance within the range measured for relatively neutral gas in the source  \citep{Yaqoob:1993aa,Schurch:2003aa,Nandra:2007aa}. We are also motivated to do this to limit the influence of the degeneracy between the black hole spin and the Fe abundance  \citep{Reynolds:2012ab}. As shown in Table~\ref{tab:taModMot}, the inclusion of the inner disk reflection component is motivated by a reduction in $\chi^2/\nu$ by $\Delta \chi^2 / \Delta \nu = -1060 /-7$. 

\begin{figure*}
 \centering
\includegraphics[width=0.70\textwidth, angle=270]{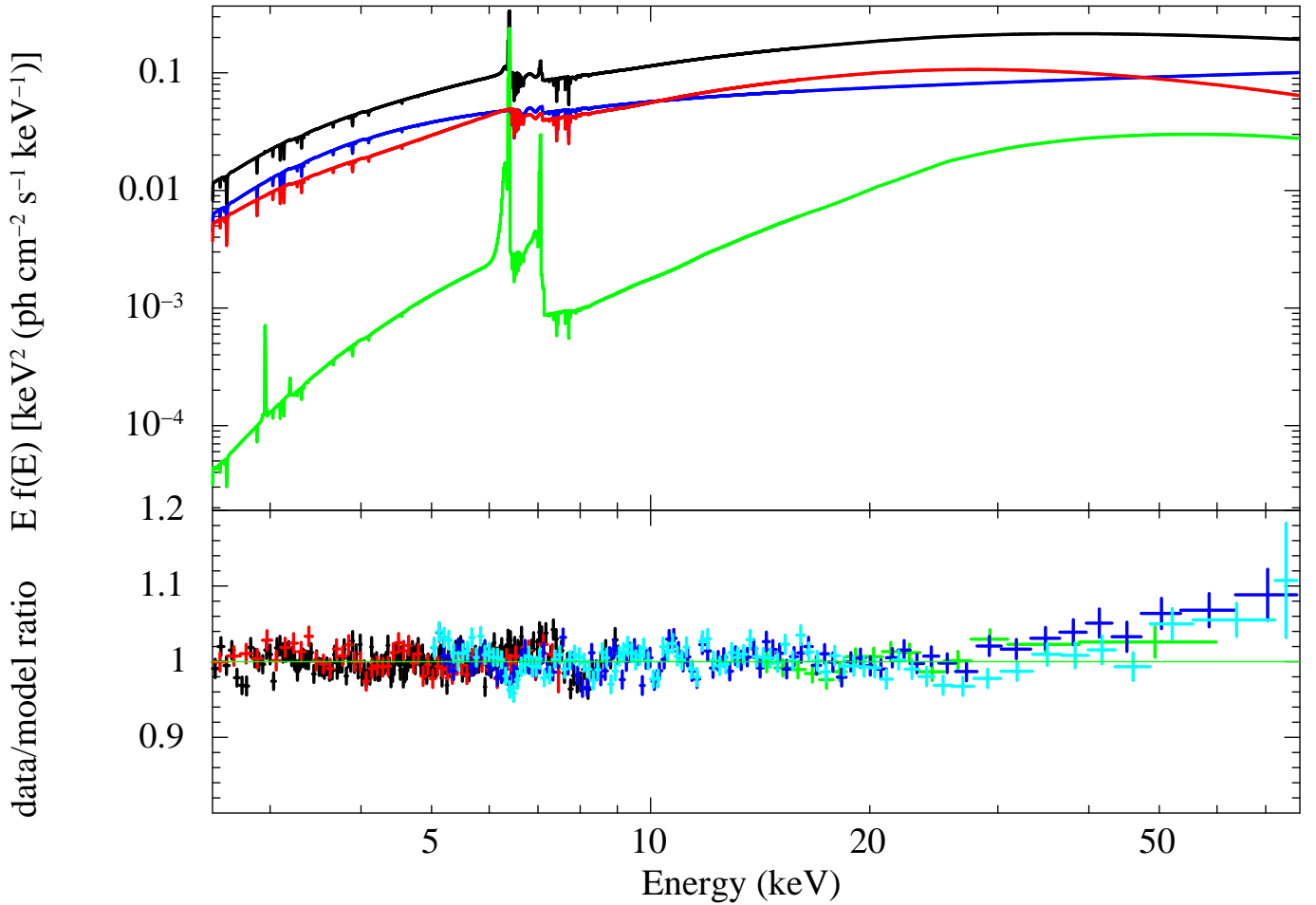}
 \caption{ \textit{Top Panel:} Model plotted in $Ef(E)$. The components are, from top to bottom, the total model (black), the cut-off power-law (dark blue), the inner disk reflection component (red), and the distant reflector component (green). \textit{Bottom Panel:} The data/model ratios for XIS-FI (black), XIS-1 (red), PIN (green), FPMA (dark blue), and FPMB (light blue). The green line indicates a data/model ratio of unity. For plotting purposes only, spectral bins have either a significance of $80\sigma$ or a maximum of $80$ spectral channels. (A color version of this figure is available in the online journal.) \label{fig:eemodratall}}
\end{figure*}

 In summary, our model has the functional form {\tt (Tbabs)*((partcov*zTbabs)*WA1*WA2*(xillver + highecut*zpowerlw + relconv*xillver))}. Best-fit parameter values are shown in Table~\ref{tab:tabfvZPO}, and the best-fit model and data/model ratios are shown in Figure~\ref{fig:eemodratall}. 

\begin{deluxetable*}{llc}
\tablecaption{ Time-averaged spectral analysis best-fit values for the inner disk reflection model    \label{tab:tabfvZPO}}
\tablehead{\colhead{Component} & \colhead{Parameter (Units)} & \colhead{Values} \\ \colhead{ } & \colhead{ } & \colhead{ } }
\startdata
partial covering  & $N_H$ ($\pcmsq$) & $(15.4\pm0.6)\times 10^{22}$  \\
({\tt partcov*zTbabs}) & $f_{cov}$& $0.92\pm0.01$  \\
warm absorber 1 & $N_H$ ($\pcmsq$) & $2.8_{-0.3}^{+0.5}\times 10^{22}$  \\
({\tt XSTAR} grid) & $\mathrm{log(}\xi/\ergcmps$) & $2.5_{-0.0p}^{+0.1}$     \\
warm absorber 2 & $N_H$ ($\pcmsq$) & $2.4_{-0.6}^{+0.5}\times 10^{22}$  \\
({\tt XSTAR} grid) & $\mathrm{log(}\xi/\ergcmps$) & $3.3\pm0.1$     \\
distant reflector & $\AFe$  & $5.0_{-0.1}^{+0.0p}$  \\
({\tt xillver}) & $K_{DRC}$ ($\phpcmsqps$) & $(1.7\pm0.1)\times 10^{-4}$  \\
	cut-off power-law & $\Gamma$ & $1.75_{-0.02}^{+0.01}$   \\
                   & $E_{cut} ($\keV$) $ & $1000 (f)$  \\
& $K_{PLC}$ ($\phpcmsqps$) & $3.7_{-0.3}^{+0.4}\times 10^{-2}$ \\
inner disk reflection& $a$  & $0.98\pm0.01$   \\
({\tt relconvf*xillver})& $q_{1}$  & $10.0_{-0.4}^{+0.0p}$  \\
& $q_{2}$  & $2.9\pm0.1$  \\
& $r_{br}$ ($\rg$) & $3.3\pm0.1$  \\
  & $r_{in}$  & $r_{ISCO}$ (f)   \\
   & $r_{out}$ ($\rg$) & $400$ (f)   \\
 &  $\mathrm{log(}\xi/\ergcmps$) & $3.07\pm0.02$  \\
 & $i$ ($\deg$) & $18_{-0p}^{+2}$ \\
 & $K_{IDR}$ ($\phpcmsqps$) & $(1.3\pm0.1)\times 10^{-6}$  \\
 cross-normalization\tablenotemark{a} & XIS-1 & $0.96\pm0.01$  \\
& PIN & $1.26\pm0.01$  \\
 & FPMA& $1.04\pm0.01$  \\
 & FPMB & $1.08\pm0.01$  \\
flux, absorbed\tablenotemark{b} & $F_{2-10}$ ($\ergpcmsqps$)& $1.50$$\times 10^{-10}$  \\
 &  &   \\
fit & $\chi^{2}/\nu$ & $4453/4101$$(1.09)$   \\

\enddata
\tablecomments{Statistical errors are quoted to $90\%$ confidence for one interesting parameter. The redshifts of all components of the source are fixed to the systemic value, $z = 0.00332$. The Fe abundances of the IDR and distant reflector are linked together in the fitting. The letter `p' is included in confidence limits that have pegged at a limiting value allowed in the model.} 
\tablenotetext{a}{Cross-normalizations are relative to XIS-FI.}
\tablenotetext{b}{The model calculated $2-10 \keV$ flux is quoted for XIS-FI.}
\end{deluxetable*}

\begin{figure}
 \centering
\includegraphics[width=0.49\textwidth]{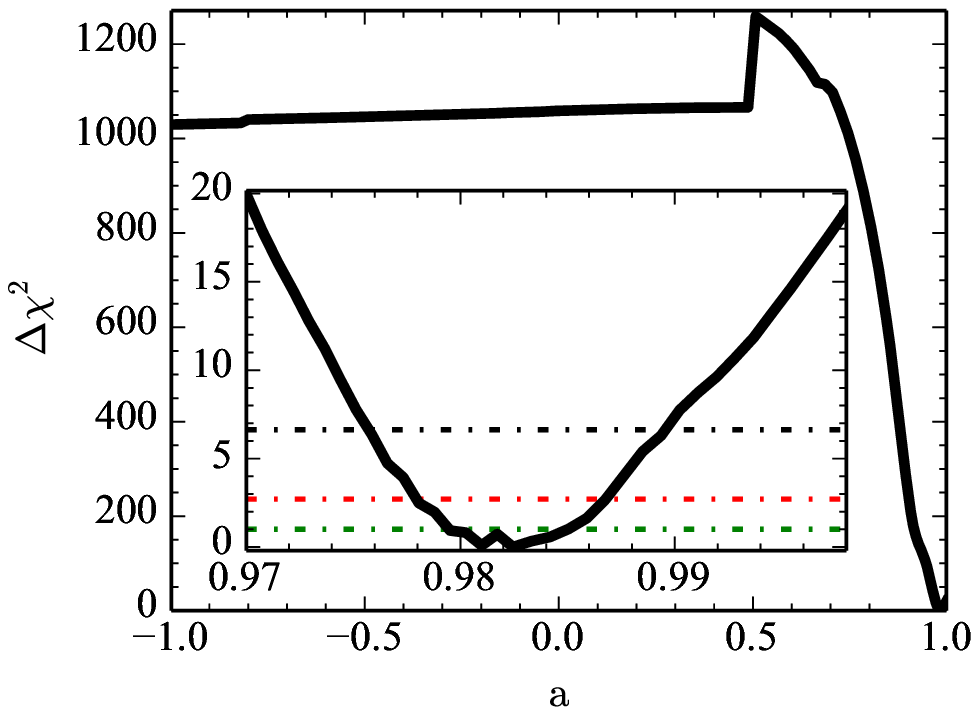}
 \caption{$\Delta \chi^2$ versus the dimensionless spin parameter illustrating our statistical spin constraint. The inset shows the range $a>0.97$. $68\%$, $90\%$, and $99\%$ confidence levels for one interesting parameter are indicated with dashed lines, which correspond to $\Delta \chi^2$ values of 1.00, 2.71, and 6.63, respectively. (A color version of this figure is available in the online journal.) \label{fig:contoursA}}
\end{figure}

 Applying an inner disk reflection model, we find strong evidence for relativistic reflection and a preliminary SMBH spin constraint $a=0.98\pm0.01$ as shown in Figure~\ref{fig:contoursA}. If we calculate the reflection fraction, $R$, as the ratio of the inner disk reflector flux to the power-law component flux in the $20-40 \keV$ band, we find that $R=1.3\pm0.2$. This reflection fraction is relatively low given the near-maximal spin and relatively steep inner emissivity index, $q_1=10.0_{-0.4}^{+0.0p}$, with respect to the maximum expected reflection fraction in the lamp post geometry \citep{Dauser:2014aa}. We discuss this further in \S\ref{sec:d}. 
 
\begin{figure*}
 \centering
\includegraphics[width=0.495\textwidth]{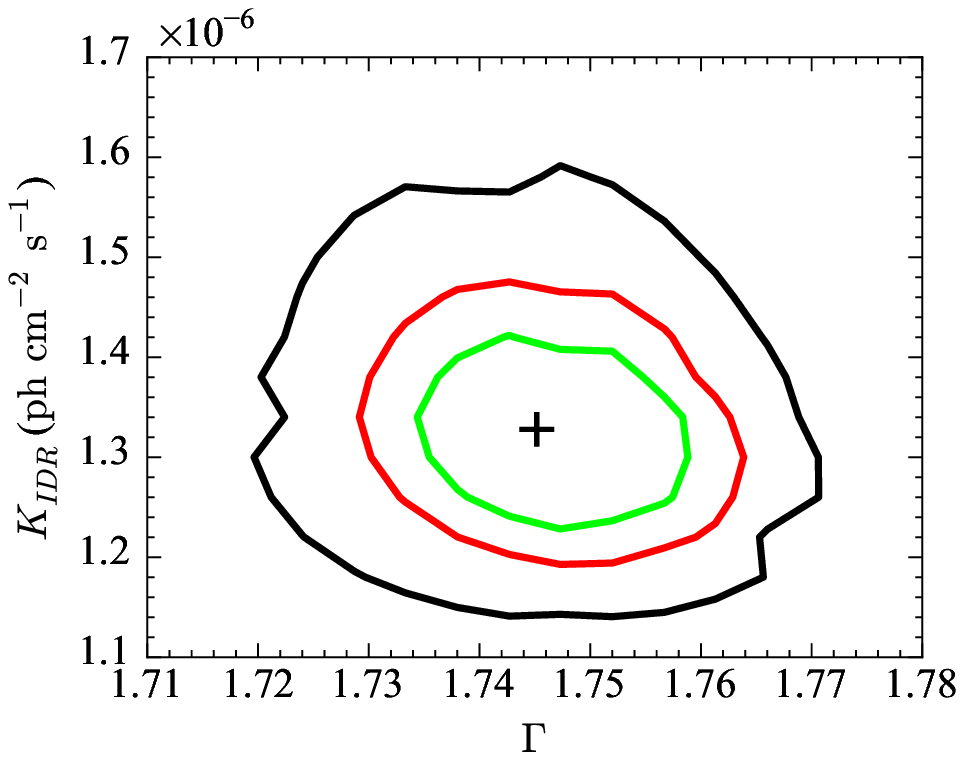}
\includegraphics[width=0.495\textwidth]{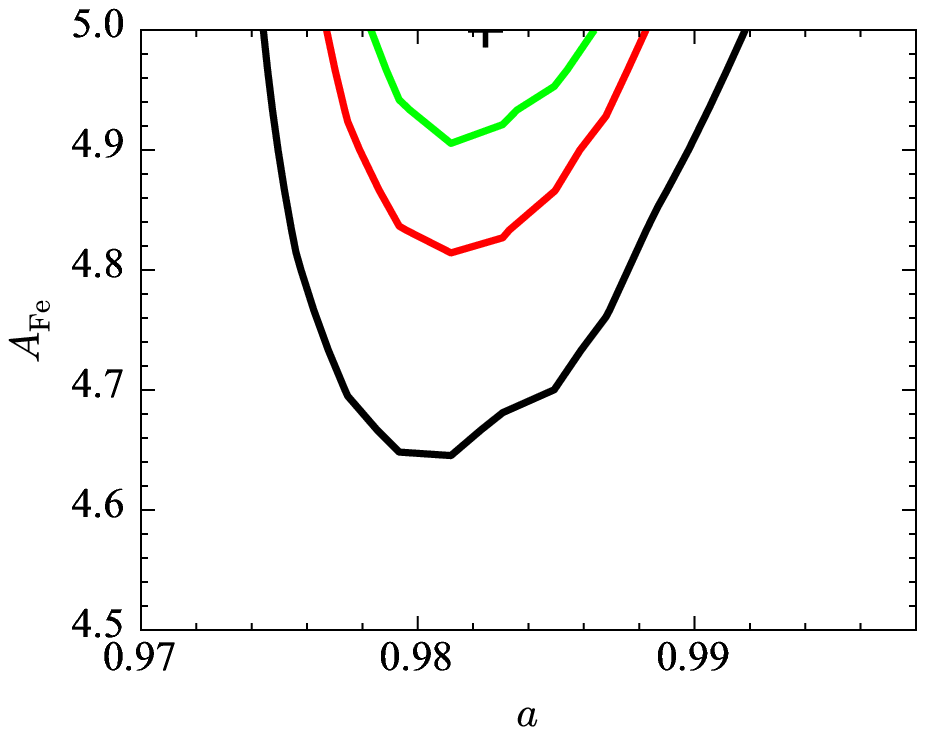}
\includegraphics[width=0.495\textwidth]{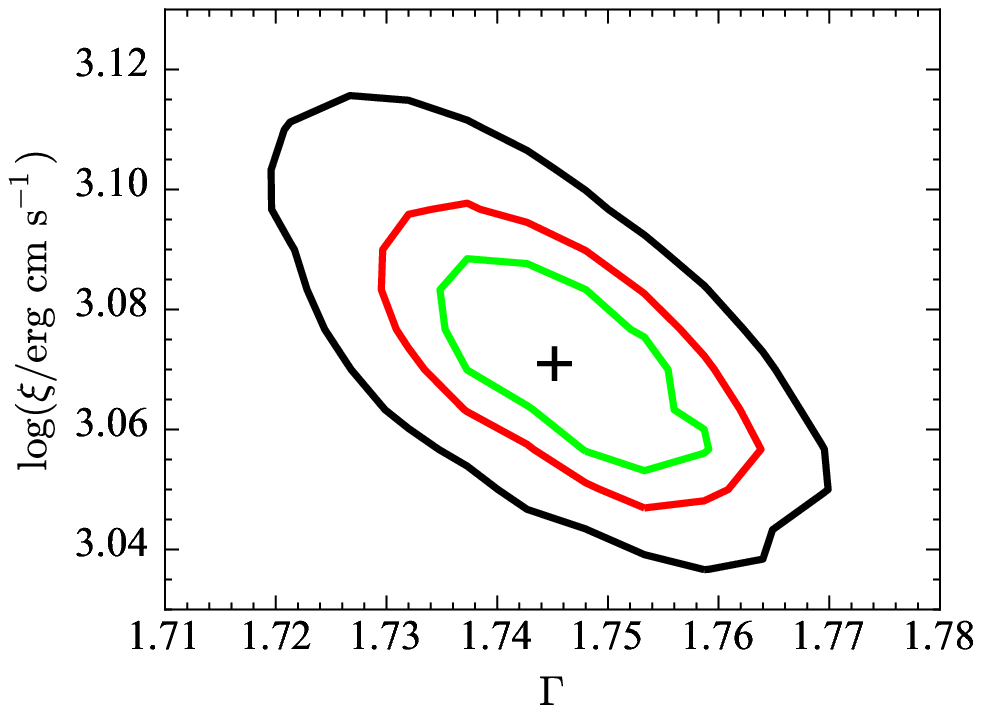}
\includegraphics[width=0.495\textwidth]{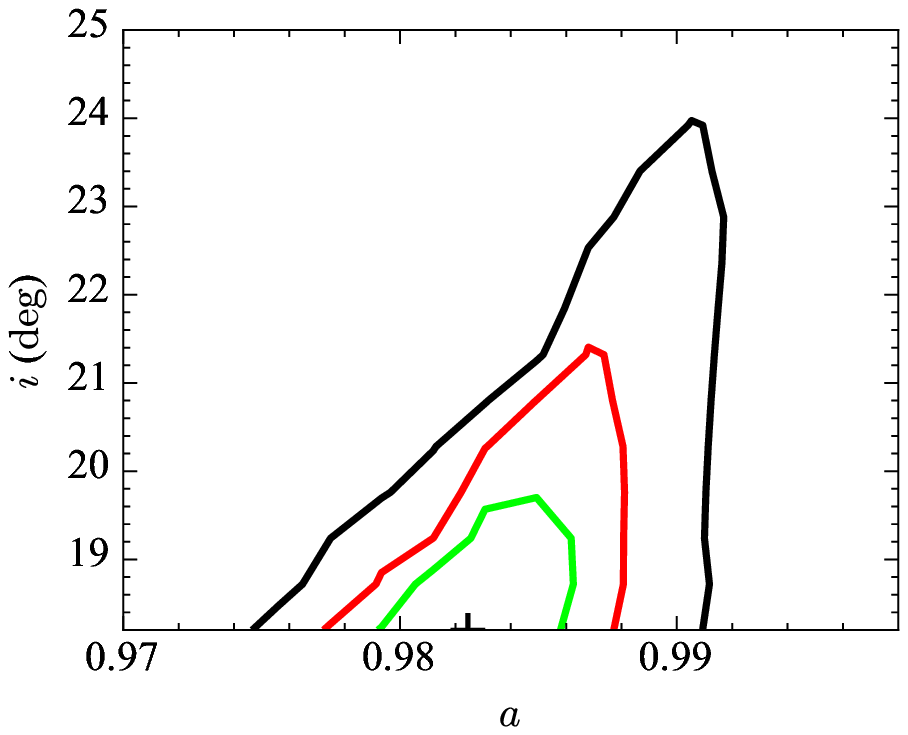}
\includegraphics[width=0.495\textwidth]{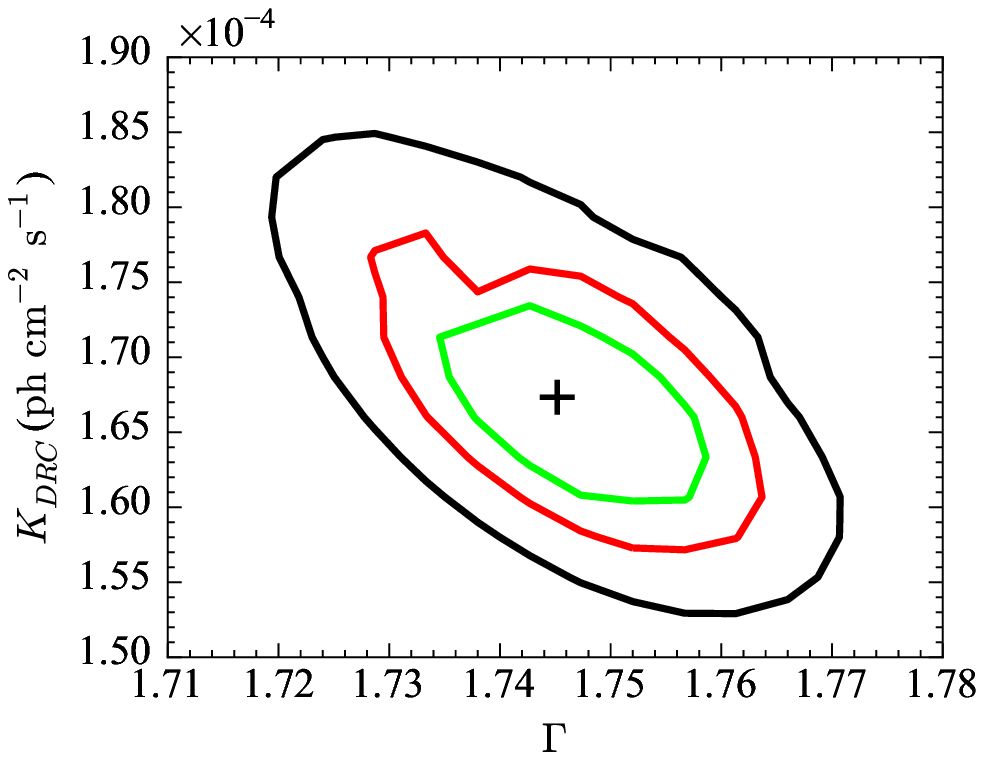}
\includegraphics[width=0.495\textwidth]{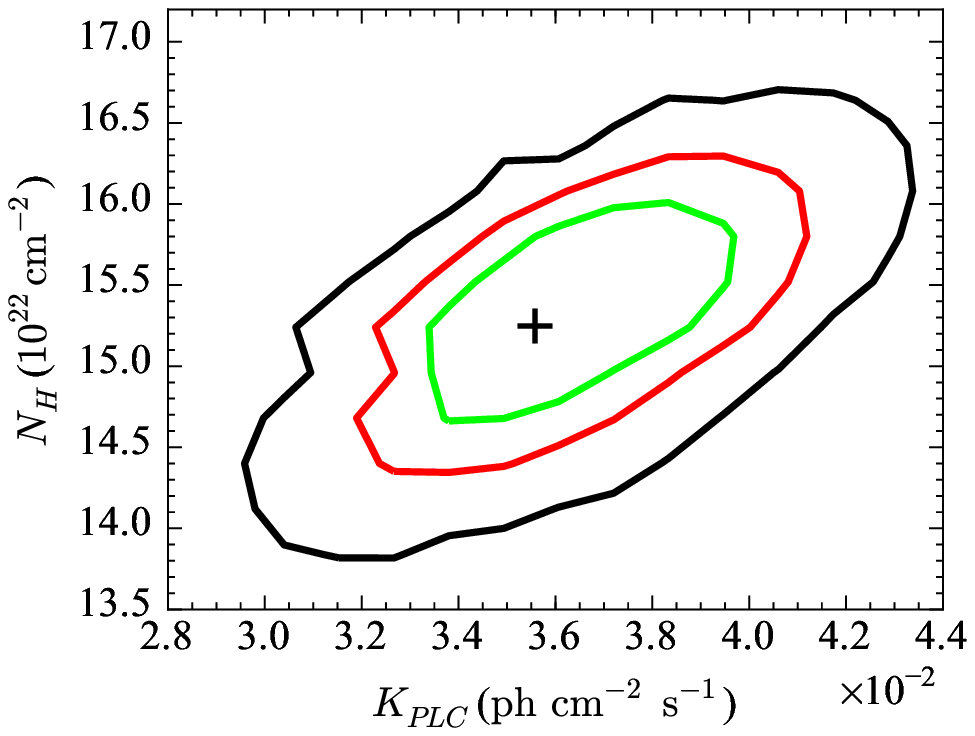}
 \caption{Confidence contours for the inner disk reflection time-averaged model. Confidence contours indicate $99 \%$ (black), $90 \%$ (red), and $68 \%$ (green) confidence intervals in two interesting parameters, which correspond to $\Delta \chi^2$ values of 9.21, 4.61, and 2.30, respectively.  Confidence contours are shown for the plane of the PLC $\Gamma$ and the IDR normalization (\textit{Top Left}), the plane of $a$ and $\mathrm{A_{Fe}}$ (\textit{Top Right}), the plane of the PLC $\Gamma$ and the IDR $\xi$ (\textit{Middle Left}), the plane of the IDR $a$ and inclination (\textit{Middle Right}), the plane of the PLC $\Gamma$ and the distant reflector normalization (\textit{Bottom Left}), and the plane of the PLC normalization and the partial covering absorber $N_H$ (\textit{Bottom Right}).    (A color version of this figure is available in the online journal.) \label{fig:contoursCor}}
\end{figure*}

We find a partial covering fraction that is consistent with a scattered fraction. Parameters largely do not show degeneracies as illustrated with confidence intervals in Figure~\ref{fig:contoursCor}. We do find a degeneracy between the inner disk component spin and inclination although we find values for these parameters that are consistent with previous studies as discussed in \S\ref{sec:d}. For the cut-off power-law component,  we find only a lower limit $E_{cut}>600 \keV$ if we allow the cut-off energy to fit freely. For simplicity, we fix $E_{cut}=1000 \keV$ in the best-fit model and find $\Gamma =1.75_{-0.02}^{+0.01}$.

\begin{figure}
 \centering
\includegraphics[width=\fwb, angle=270]{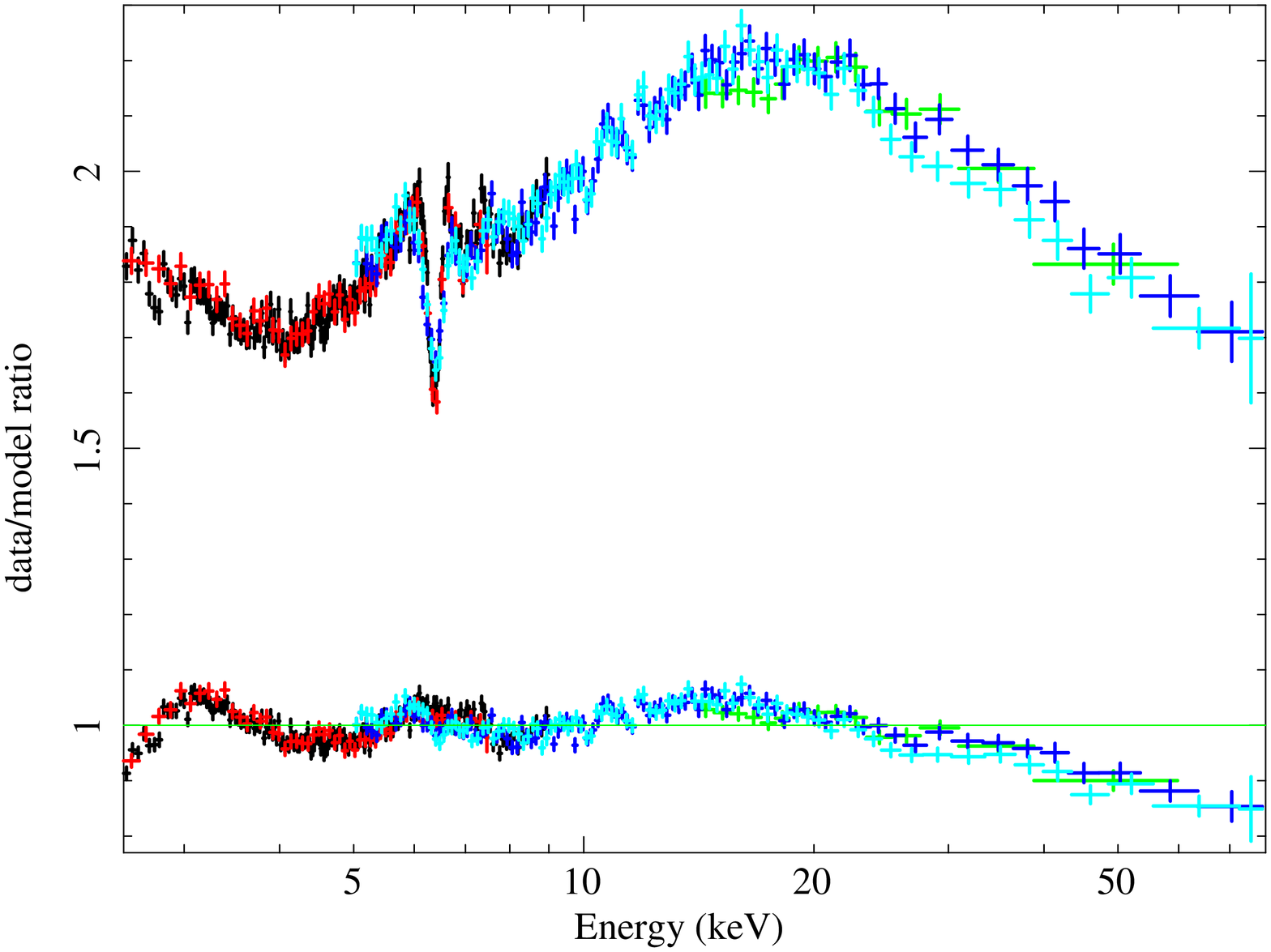}
 \caption{Data/model ratios are shown after starting with the best-fit model and then removing the inner disk reflection component without refitting (ratios at top) to illustrate the spectral features this component accounts for. Best-fit ratios after refitting with $\Gamma=1.75$ fixed are shown on the bottom. Spectra are rebinned strictly for plotting purposes to have bins either with a signal-to-noise of $80\sigma$ or a maximum of 80 spectral channels. Also, the green line indicates a data/model ratio of unity.   (A color version of this figure is available in the online journal.)    \label{fig:tasnADR}}
\end{figure}

To illustrate the spectral features that are modeled by the inner disk reflection component, we remove this component from the best-fit model giving the data/model ratios shown in Figure~\ref{fig:tasnADR}. We then refit this model, which gives a significantly worse fit than the best-fit model ($\Delta \chi^2 / \Delta \nu = +1060 /+7$). This increase in $\chi^2$ is driven by residuals in the Fe K-band and the energy range $E\gtrsim 10 \keV$ identical to those motivating its inclusion.

We note that non-unity ratios remain at the $<10\%$ level in the best-fit data/model ratios at energies $E\gtrsim30 \keV$, as $\chi^2/\nu = 1304/1219 \, (1.07)$ ignoring data below $30 \keV$.  We attempt to account for these non-unity ratios with a phenomenological hard power-law component, which have been observed in radio-loud AGNs (e.g., 3C 273; \citealt{Grandi:2004aa}), in order to test whether this may be emission from the dim, extended radio jet in the source \citep{Mundell:2003aa,Wang:2011ac}. The hard power-law component fits with $\Gamma_{hard}=1.23_{-0.10}^{+0.05}$, and the coronal power-law $\Gamma$ increases to $\Gamma=1.83_{-0.02}^{+0.03}$.  This component gives a small but significant reduction in the fit statistic of $\Delta \chi^2 / \Delta \nu = -39 / -2$. The presence of such a hard power-law component, which could perhaps correspond to extended jet emission, would be interesting although we caution that the component is mainly significant in an energy band where there is more uncertainty in the \textit{NuSTAR} calibration. 

In attempt to constrain the outer extent of the corona, we try replacing the \texttt{relconv} component in the best-fit model with a relativistic convolution component with a twice broken power-law emissivity profile,  {\tt kdblur3} \citep{Wilkins:2011aa}. We allow the inner disk radius to fit freely because the spin is fixed at $a=0.998$. The outer index is fixed to $q_3=3$ as expected from basic geometry at radii in the disk where $r\gg h$. We limit the outer break radius to values $r_{br,2}>10 \rg$, and we find $r_{br,2}=10_{-0p}^{+6} \rg$. We also find $r_{in}=1.5\pm0.1 \rg$, $q_{1}=10.0_{-0.4}^{+0.0p}$, $r_{br,1}=3.5\pm0.1 \rg$, and $q_{2}=2.5_{-0.2}^{+0.3}$. This model provides only a slightly better fit than the best-fit model ($\Delta \chi^2 / \Delta \nu = -4 / -1$), so a twice-broken power-law emissivity profile is not strongly warranted statistically.

If we apply a lamp post model, \texttt{relconv\_lp} \citep{Dauser:2013aa}, in place of \texttt{relconv} in the best-fit model, we find a lamp post height $h=1.3_{-0.0p}^{+0.1} \rg$ and a reflection fraction consistent with the best-fit model ($R=1.2\pm0.1$) although we also find a significantly worse fit ($\Delta\chi^2/\Delta\nu = +493/+2$ relative to the best-fit model). We note that the worse fit with \texttt{relconv\_lp} results from the inner disk emissivity profile being determined solely with the lamp post height compared to the broken power-law profile used in \texttt{relconv}. 

We attempt to add a second lamp post component by including a second inner disk component because coronae could be extended, such as in a scenario where a jet produces the coronal emission or a corona has multi-site activity. We fix all of its parameters to those of the first inner disk component except for the normalization and lamp post height. We find a similar reflection fraction and an improved fit relative to the single lamp post model ($\Delta\chi^2/\Delta\nu = -240/-2$) where the second lamp post component has height $h=17\pm3\rg$ and low flux relative to the first lamp post component, which again fits with a low height $h=1.3_{-0.0p}^{+0.3} \rg$. This fit along with our best-fit model indicate that a model more complicated than a single point source on the spin axis provides a statistically favored description of the reflection spectrum. This fit also suggests that the corona may have an extended component.

In attempt to test the extended corona scenario self-consistently, we apply the {\tt relconv\_LP\_ext} model \citep{Dauser:2013aa}. We model the disk-illuminating source as an outflow on the spin axis with outflow velocity $v_{out}$, base height $h_{base}$, and maximum height $h_{top}$. We fix $h_{base}$ to the event horizon radius as motivated by the dual lamp post model described above. We find $v_{out}/c = 0.01_{-0.01p}^{+0.08}$ and  $h_{top} = 1.3_{-0.0p}^{+0.1} \rg$. This model (which gives $\Delta\chi^2/\Delta\nu = +477/+1$ relative to the best-fit model) is consistent with the single lamp post model presented above within errors. We find a better fit with the dual lamp post model than with this extended corona model, which may result from describing the corona with two physically discrete illuminating components rather than a continuous illuminating source. 

 We try applying the {\tt relxill}  model in place of the IDR component in the best-fit model to more accurately account for the emission angle as a function of radius. The inner disk reflection models we employ in the best-fit model assume a single emission angle for the inner accretion disk emission and also assume an isotropic emission distribution. While these are valid assumptions for constraining properties of relativistic reflection, the emission angle of photons leaving the disk is not constant and depends upon radius, especially in the innermost regions of the disk as a consequence of relativistic light-bending \citep{Garcia:2014aa}.   We find parameter values consistent within errors with those for the best-fit model including a reflection fraction in the $20-40 \keV$ band $R=1.3\pm0.2$ and a similar fit to the best-fit model ($\Delta\chi^2/\Delta\nu = -5/+0$). Consequently, the assumption of a single emission angle for the inner accretion disk does not introduce significant systematic error into our results. 

\subsubsection{Absorption-dominated model}\label{sec:taaao}

\begin{figure}
 \centering
\includegraphics[width=0.33\textwidth, angle=270]{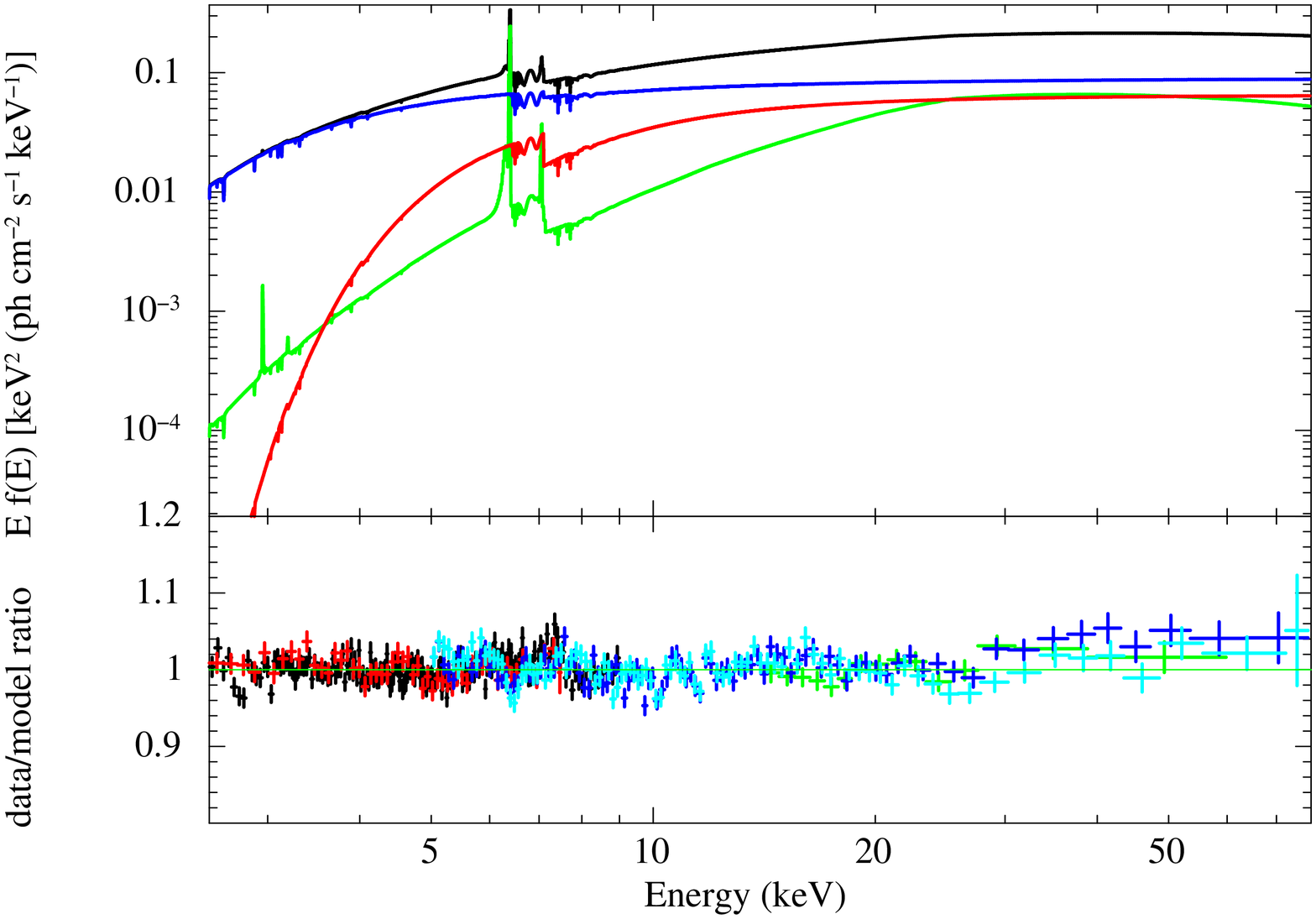}
 \caption{ \textit{Top Panel:} Absorption-dominated model plotted in $Ef(E)$. The broadband components are, from top to bottom, the total model (black), the coronal power-law component that is absorbed by PC2 (red), the power-law component that is not absorbed by PC2 (blue), and the distant reflector component (green). \textit{Bottom Panel:} The data/model ratios for XIS-FI (black), XIS-1 (red), PIN (green), FPMA (dark blue), and FPMB (light blue). The green line indicates a data/model ratio of unity. For plotting purposes only, spectral bins have either a significance of $80\sigma$ or a maximum of $80$ spectral channels. (A color version of this figure is available in the online journal.) \label{fig:eemodratao}}
\end{figure}

We investigate an absorption-dominated model and develop a model that has the functional form \texttt{(Tbabs) * (WA1 * WA2 * (partcov*zTbabs) * (partcov*zTbabs) * (xillver + zpowerlw))}. We find a best-fit model shown in Table~\ref{tab:tabfvao} and Figure~\ref{fig:eemodratao}. This model provides a statistically less satisfactory description of the time-averaged data ($\Delta\chi^2/\Delta\nu = +86/+5$) relative to the best-fit model.

\begin{deluxetable}{llc}
\tablecaption{{Absorption-dominated model time-averaged best-fit values} \label{tab:tabfvao}}
\tablehead{\colhead{Component} & \colhead{Parameter (Units)} & \colhead{Value} \\ \colhead{ } & \colhead{ } & \colhead{ } }
\startdata
PC1   & $N_H$ ($\pcmsq$) &   ($13.3\pm0.9)\times 10^{22}$ \\
  & $f_{cov}$& $0.94\pm0.01$   \\
PC2 & $N_H$ ($\pcmsq$) & $56_{-4}^{+3}\times 10^{22}$  \\
  & $f_{cov}$& $0.42\pm0.02$   \\
  Warm absorber 1 & $N_H$ ($\pcmsq$) & $1.6_{-0.3}^{+0.5}\times 10^{22}$  \\
   & $\mathrm{log(}\xi/\ergcmps$) &  $2.52_{-0.02p}^{+0.07}$    \\
Warm absorber 2 & $N_H$ ($\pcmsq$) & $5.9_{-0.4}^{+0.5}\times 10^{22}$  \\
   & $\mathrm{log(}\xi/\ergcmps$) &  $3.46\pm0.03$    \\
DRC & $\AFe$  & $1.5_{-0.1}^{+0.2}$  \\
  & $K_{DRC}$ ($\phpcmsqps$) & $(4.3\pm0.2)\times 10^{-4}$  \\
 & $\mathrm{log(}\xi/\ergcmps$) & $0$ (f)  \\
PLC  & $\Gamma$ & $1.92\pm0.01$   \\
                   & $E_{cut} ($\keV$) $ & $1000 (f)$  \\
& $K_{PLC}$ ($\phpcmsqps$) & $0.117_{-0.003}^{+0.002}$  \\
 &  &   \\
Fit & $\chi^{2}/\nu$ & $4539/4106 \, (1.11)$   \\
\enddata
\tablecomments{Statistical errors are quoted to the $90\%$ confidence level. Instrument cross-normalizations and model-predicted fluxes are consistent with those shown in Table~\ref{tab:tabfvZPO}. The redshifts of all components of the source are fixed to the systemic value, $z = 0.00332$. PC1 and PC2 indicate partial-covering absorbers 1 and 2, DRC indicates distant reflection component, and PLC indicates power-law component. } 
\end{deluxetable}

The absorption-dominated model also has positive residuals at the $<10\%$ level at energies $E>30 \keV$, so we also attempt adding a hard power-law component to it. We find a similar hard power-law component as for the inner disk reflection model with $\Gamma_{hard}=1.25_{-0.07}^{+0.08}$ can account for the high energy non-unity ratios. We also find a moderate reduction in the fit statistic of $\Delta \chi^2 / \Delta \nu = -24 / -2$.

\subsection{Time-Resolved Analysis}\label{sec:tra}

To perform the time-resolved analysis, we produce strictly simultaneous time-resolved data products (i.e., the data products are time-filtered to only include times when the two observatories were simultaneously viewing NGC 4151)  in seven time-intervals as shown in Figure~\ref{fig:hrnusu}. Intervals are chosen to maximize S/N and have approximately constant hardness ratios. We omit making PIN time-resolved products due to \textit{NuSTAR's} superior sensitivity at $E>10 \keV$. Information for the time-resolved data is shown in Table~\ref{tab:trinfDT}. We note that  similar time-resolved analyses have been performed on joint \textit{NuSTAR} observations with \textit{Suzaku} or \textit{XMM-Newton} of the Seyfert galaxies IC 4329A \citep{Brenneman:2014ab}, NGC 1365 \citep{Walton:2014aa}, and MCG 6-30-15 \citep{Marinucci:2014ab}.

 \begin{deluxetable}{ccccccc}
   \tablecaption{Exposure times and background subtracted counts for the \textit{Suzaku} and \textit{NuSTAR} time-resolved data  \label{tab:trinfDT}} 
 \tablehead{   & \multicolumn{2}{c}{XIS-FI} & \colhead{XIS-1} &  \multicolumn{2}{c}{FPMA} & \colhead{FPMB}   \\
 \colhead{Int.}  & \colhead{Time} & \colhead{Counts} & \colhead{Counts}  &  \colhead{Time} & \colhead{Counts} &  \colhead{Counts} \\
  \colhead{}  & \colhead{($\s$)} & & &  \colhead{($\s$)} & &
 }
      \startdata
    1     & 2776  & 18211 & 7245  & 2959  & 13225 & 12946 \\
    2     & 19535 & 137709 & 55138 & 18680 & 98549 & 90109 \\
    3     & 6920  & 40678 & 15810 & 6319  & 31340 & 30462 \\
    4     & 24474 & 127483 & 49964 & 23550 & 108053 & 103254 \\
    5     & 7640  & 48024 & 18871 & 7318  & 39132 & 37702 \\
    6     & 2423  & 17205 & 7043  & 2906  & 16372 & 15494 \\
    7     & 6055  & 35980 & 14411 & 7120  & 34199 & 32830 \\
\enddata
\tablecomments{Counts correspond to the $2.5-9 \keV$ energy bandpass for XIS-FI, $2.5-7.5 \keV$ for XIS-1, and $5-79 \keV$ for FPMA and FPMB. The relatively low exposure times for the interval 6 data result from relatively poor overlap between the \textit{NuSTAR} and \textit{Suzaku} observations during this interval.}
\end{deluxetable}

\begin{figure*}
 \centering
\includegraphics[width=0.37\textwidth, angle=270]{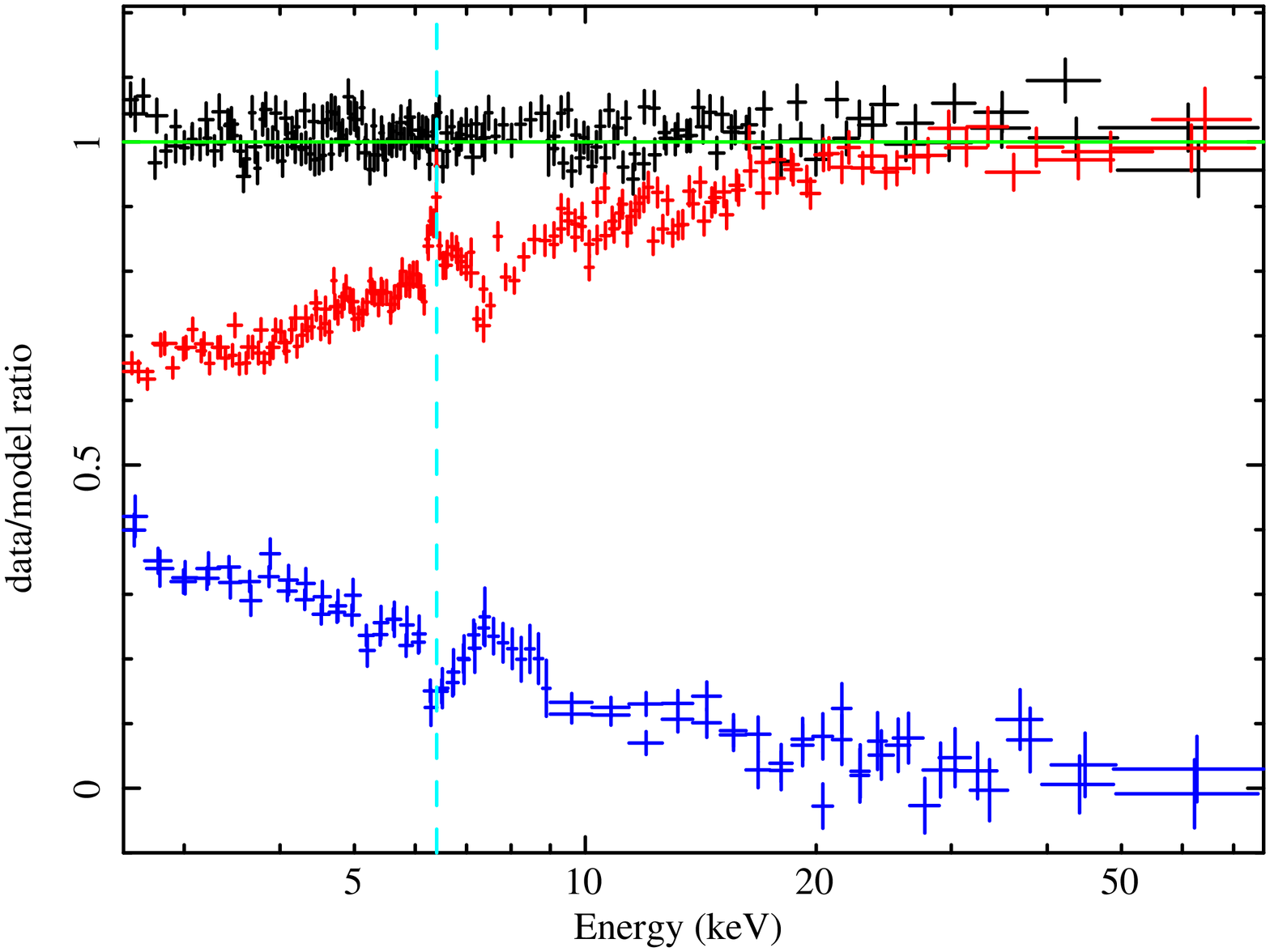}
\includegraphics[width=0.37\textwidth, angle=270]{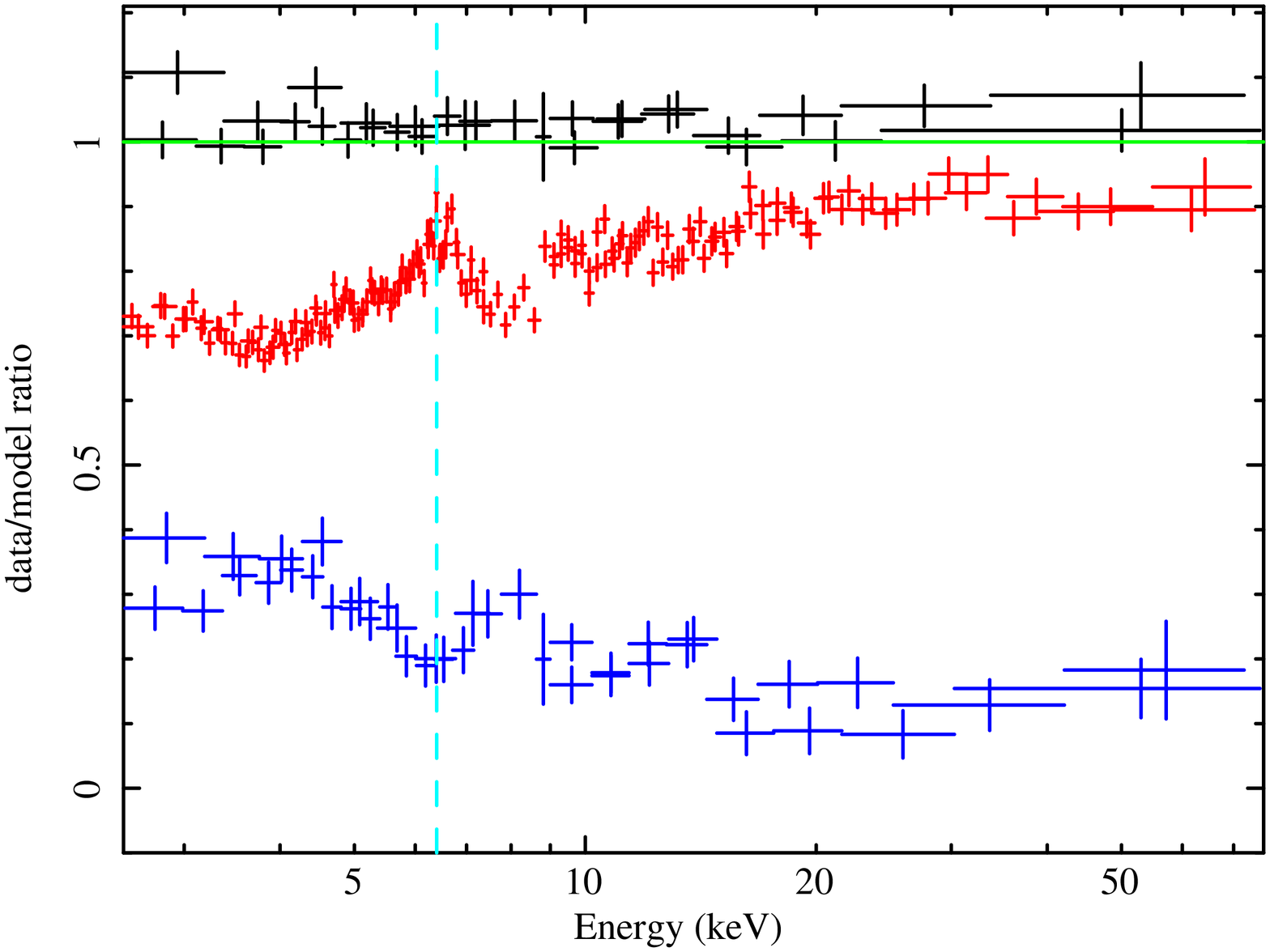}
 \caption{ \textit{Left:} The ratios of the interval 2 (black, top), interval 4 (red, middle), and difference (blue, bottom) spectra to the best-fit model to interval 2. \textit{Right}: The ratios of the interval 6 (black, top), interval 4 (red, middle), and difference (blue, bottom) spectra  to the best-fit model to interval 6. Notice the narrow dip in both difference spectra at the energy of neutral Fe K$\alpha$ ($6.4 \keV$ as indicated with a dashed, light blue line). XIS-FI, XIS-1, FPMA, and FPMB spectra are shown although only \textit{NuSTAR} data is shown above $9 \keV$ and only \textit{Suzaku} data is below $9 \keV$ for clarity.  Spectra have been rebinned for plotting purposes only.   The solid green line indicates a data/model ratio of unity. (A color version of this figure is available in the online journal.) \label{fig:dsall}}
\end{figure*}

To first assess the spectral variability in a model independent fashion, we analyze difference spectra between high flux states (intervals 2 and 6) and a low flux state (interval 4) as shown in Figure~\ref{fig:dsall}. The difference spectra peak at  $E\sim 3 \keV$ and gradually decrease with increasing energy. Additionally, the flux of the narrow, neutral Fe K$\alpha$ line at $6.4 \keV$ varies less than the continuum. 

To fully assess the spectral variability, we perform a joint fit of the time-resolved spectra using a strategy similar to our time-averaged analysis. We initially set the parameter values to their time-averaged model values (which are shown in Tables~\ref{tab:tabfvZPO} and \ref{tab:tabfvao}).  We require many parameters to fit freely to the same value across all intervals on physical grounds. We assume the instrument cross-normalizations are constant during the observations. The distant reflection component, which likely arises from material at distances in the range $r\sim7$ light-days (which corresponds to the approximate distance to the optical broad-line region in the source; \citealt{Bentz:2006aa}) to $r\sim 4.2\times10^4$ light-days (which is the upper-limit on the size of a dusty torus in the source; \citealt{Radomski:2003aa}), likely does not vary on the time-scale of the observations as also suggested by the RMS $F_{var}$ of the data. The Fe abundance and inclination of the inner accretion disk also can  be assumed to be constant during the observations. Similarly,  the black hole spin can be safely assumed to be constant, as SMBH angular momentum evolves only on cosmic time-scales from accretion and/or mergers (e.g., \citealt{Berti:2008aa}). We additionally fix the WA2 ionization parameter to its time-averaged best-fit value.

\subsubsection{Inner disk reflection model}\label{sec:traidr}

We systematically assess which parameters should be variable and constant in the time-resolved analysis by carrying out fits for all combinations of variable parameters for the following seven parameters: neutral absorber $N_H$, $f_{cov}$, DRC normalization ($K_{DRC}$), PLC $\Gamma$, PLC $20-40 \keV$ flux ($F_{PLC}$),  IDR $\xi$, and IDR $20-40\keV$ flux ($F_{IDR}$). We parameterize the normalizations of the PLC and IDR in terms of their $20-40 \keV$ flux for the time-resolved analysis using the \texttt{cflux} model in \texttt{XSPEC}. We allow parameters to be variable until we no longer find an improvement in the fit based on the F-test. We find a best-fit time-resolved fit with four variable parameters ($F_{PLC}$, $F_{IDR}$, $N_H$, and $\Gamma$). If we additionally try allowing $f_{cov}$ to be variable,  it provides a marginally significant improvement to the fit according to the F-test ($\Delta \chi^2 / \Delta \nu = -30 / -6$). We find that this parameter is correlated with the power-law normalization, which is likely unphysical whether the covering fraction represents a scattered fraction or a true covering fraction for a neutral absorber. Thus, we elect to keep the covering fraction constant. 

\begin{figure*}
 \centering
\includegraphics[width=0.99\textwidth]{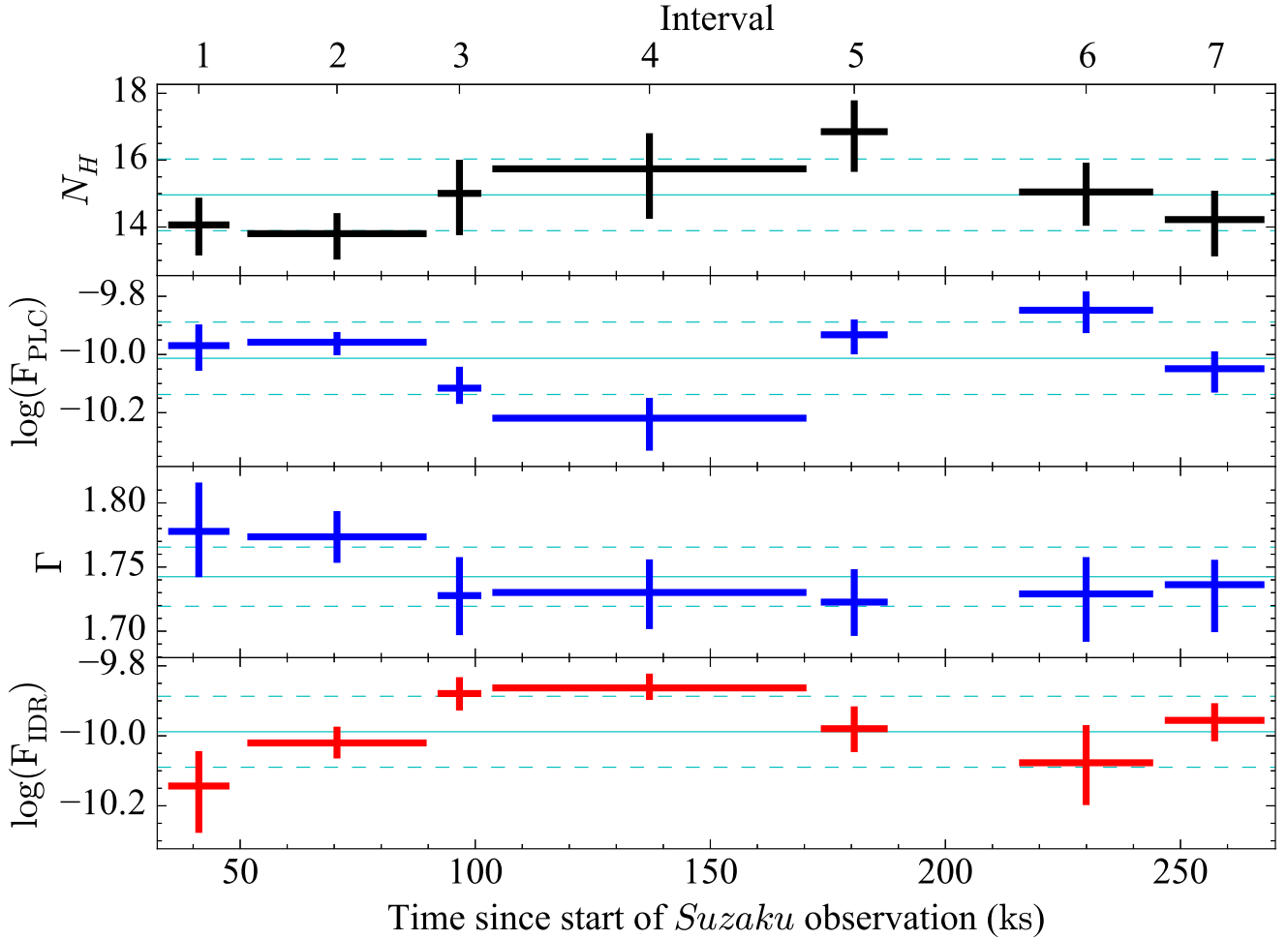}
 \caption{Time-resolved analysis best-fit values for the inner disk reflection model for the variable parameters, which are the partial covering absorber column density, power-law component flux in the $20-40 \keV$ band, power-law component photon index, and IDR $20-40 \keV$ flux. Vertical error bars indicate $90 \%$ confidence intervals, and horizontal error bars indicate the duration of the interval. Horizontal lines indicate the mean (solid) and one standard deviation (dashed). (A color version of this figure is available in the online journal.) \label{fig:ts}}
\end{figure*}

In summary, the best-fit time-resolved model values are shown in Tables~\ref{tab:tsa} and~\ref{tab:tsb}. Time-series of the parameter values allowed to vary between intervals (see Figure~\ref{fig:ts}) show that modest coronal and inner disk reflection flux variation drives the spectral variability during the observations. Note the similarity of the variation in the power-law component $20-40 \keV$ flux and the $30-79 \keV$ flux shown in Figure~\ref{fig:hrnusu}. The reflection fraction (i.e., the ratio of the inner disk reflection  and the power-law component $20-40 \keV$ fluxes) is anti-correlated with the power-law component $20-40 \keV$ flux. This anti-correlation is reasonable with respect to some of the predictions of the lamp post model for a near-maximal, prograde black hole spin.  Specifically, for small lamp post heights, the reflection fraction can increase when the lamp post moves to smaller heights due to a decrease in the continuum flux that reaches the observer \citep{Miniutti:2004aa, Dauser:2014aa}.


\begin{deluxetable*}{ccccc}
\tablecaption{Time-Resolved Best-Fit Values for Parameters Free to Fit within Each Interval for the Inner Disk Reflection Model  \label{tab:tsa}}
\tablehead{\colhead{Int.} & \colhead{$N_H$} & \colhead{$\mathrm{log}(F_{PLC}/\ergpcmsqps)$} & \colhead{$\Gamma$} & \colhead{$\mathrm{log}(F_{IDR}/\ergpcmsqps)$}\\ 
\colhead{} & \colhead{($10^{22} \pcmsq$)} & \colhead{}  & \colhead{}  & \colhead{}}
\startdata
1 & $14.1_{-1.0}^{+0.8}$ & $-9.97_{-0.09}^{+0.07}$ & $1.78\pm0.04$ & $-10.14_{-0.14}^{+0.10}$ \\
2 & $13.8_{-0.8}^{+0.6}$ & $-9.96\pm0.04$ & $1.77\pm0.02$ & $-10.02_{-0.04}^{+0.05}$ \\
3 & $15.0_{-1.2}^{+1.0}$ & $-10.12_{-0.05}^{+0.08}$ & $1.73\pm0.03$ & $-9.88\pm0.05$ \\
4 & $15.7_{-1.5}^{+1.1}$ & $-10.22_{-0.11}^{+0.07}$ & $1.73\pm0.03$ & $-9.86\pm0.04$ \\
5 & $16.9_{-1.3}^{+0.9}$  & $-9.93_{-0.07}^{+0.05}$ & $1.72_{-0.02}^{+0.03}$ & $-9.98_{-0.07}^{+0.06}$ \\
6 & $15.0_{-1.0}^{+0.9}$ & $-9.85_{-0.08}^{+0.07}$ & $1.73_{-0.04}^{+0.03}$ & $-10.08_{-0.12}^{+0.11}$ \\
7 & $14.2_{-1.1}^{+0.9}$ &  $-10.05_{-0.08}^{+0.06}$ & $1.74_{-0.04}^{+0.02}$ & $-9.96_{-0.06}^{+0.05}$ \\
Fit & $\chi^2/\nu$ & $15655/15589(1.00)$ &  &  \\
\tablecomments{$90 \%$ confidence intervals are given for the variable parameters for the inner disk reflection model, which are the partial covering absorber column density, power-law component flux in the $20-40 \keV$ band, power-law component photon index, and IDR $20-40 \keV$ flux.}
\enddata
\end{deluxetable*}

\begin{deluxetable}{ccc}
\tablecaption{Time-Resolved Best-Fit Values for Inner Disk Reflection Model Parameters Assumed to not Vary During the Observations \label{tab:tsb}}
\tablehead{\colhead{Comp.} & \colhead{Par. (units)} & \colhead{Values}\\ \colhead{ } & \colhead{ } & \colhead{ }}
\startdata
PC1 & $f_{cov}$ & $0.93\pm0.01$ \\
WA1& $N_H$ ($\pcmsq$) & $2.9_{-0.7}^{+1.0}\times10^{22}$ \\
 & $\mathrm{log(}\xi/\ergcmps$)  & $2.5(f)$  \\
WA2& $N_H$ ($\pcmsq$) & $(2.2\pm0.9)\times10^{22}$ \\
 & $\mathrm{log(}\xi/\ergcmps$)  & $3.4_{-0.1}^{+0.2}$ \\
DRC & $\mathrm{A_{Fe}}$\tablenotemark{a} & $5.0_{-0.3}^{+0.0p}$ \\
 & $K_{DRC}$ ($\phpcmsqps$) & $(1.6\pm0.2)\times 10^{-4}$ \\
IDR  & $q_{1}$ & $10.0_{-0.6}^{+0.0p}$ \\
 & $q_{2}$ & $2.8_{-0.1}^{+0.2}$ \\
 & $r_{br}$ ($\rg$) & $3.3_{-0.1}^{+0.2}$ \\
 & $a$ & $0.98\pm0.01$ \\
& $\mathrm{log(}\xi/\ergcmps$)    & $3.06_{-0.03}^{+0.02}$ \\
 & $i$ ($\deg$) & $18_{-0p}^{+5}$ \\
 cross-norm. \tablenotemark{b} & XIS-1 & $0.95\pm0.01$ \\
 & FPMA& $1.03_{-0.01}^{+0.01}$  \\
 & FPMB & $1.07_{-0.01}^{+0.01}$  \\
 Fit & $\chi^2/\nu$ & $15655/15589(1.00)$  \\
\enddata
\tablecomments{$90 \%$ confidence intervals are given. WA1 and WA2 indicate warm absorbers one and two, respectively, PC1 indicates the partial covering, neutral absorber, DRC indicates the distant reflection component, PLC indicates the power-law component, and IDR indicates the inner disk reflection component.} 
\tablenotetext{a}{The Fe abundances of the IDR and DRC components are linked together in the fitting.}
\tablenotetext{b}{Cross-normalizations are relative to XIS-FI.}
\end{deluxetable}


\subsubsection{Absorption-dominated model}\label{sec:traao}

\begin{figure*}
 \centering
\includegraphics[width=0.75\textwidth]{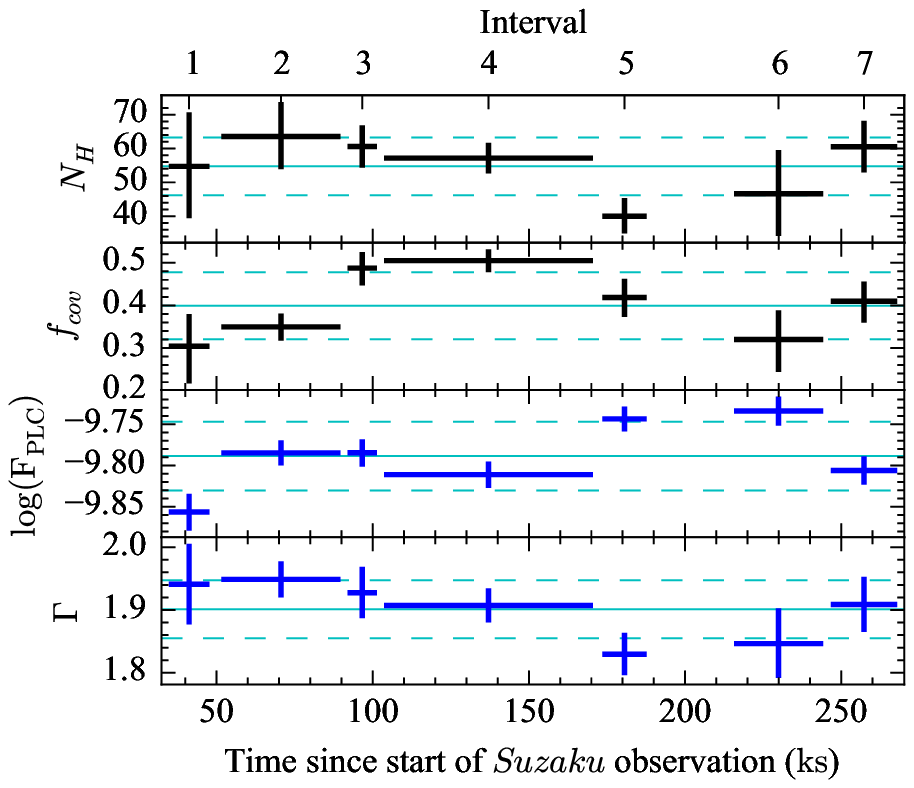}

\includegraphics[width=0.75\textwidth]{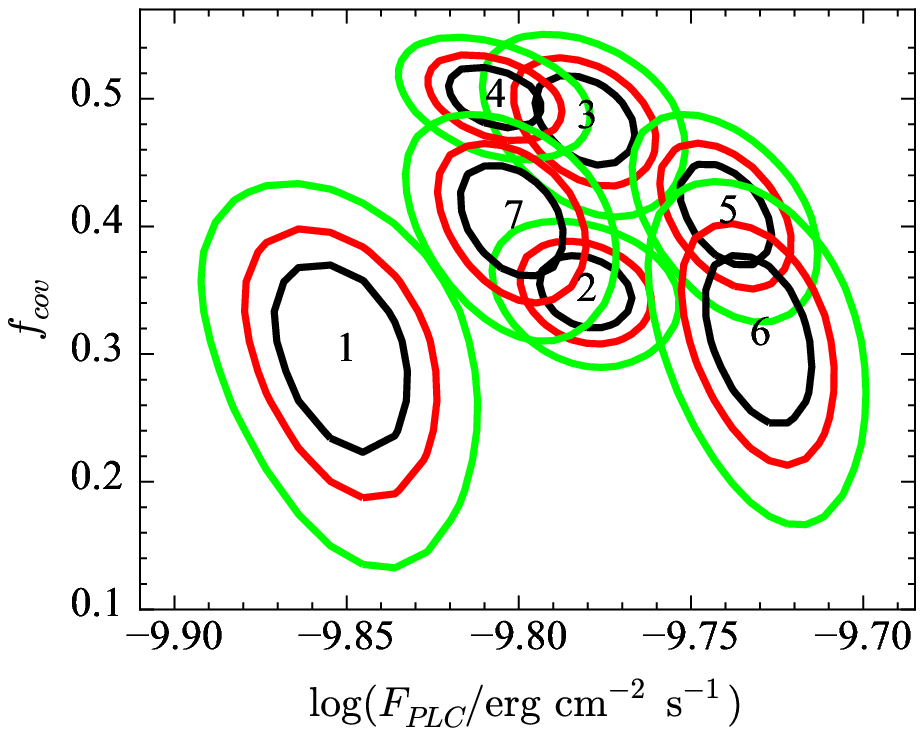}
 \caption{\textit{Top:} Best-fit values for the time-resolved analysis for the variable parameters for the absorption-dominated model, which are the PC2 covering fraction, PC2 column density, power-law component flux in the $20-40 \keV$ band, and power-law component photon index. Vertical error bars indicate 90\% confidence intervals, and horizontal error bars indicate the duration of the interval. Horizontal lines indicate the mean (solid) and one standard deviation (dashed).  \textit{Bottom:} Absorption-dominated model confidence contours for the partial covering absorber covering fraction versus the cut-off power-law flux for the time-resolved analysis for each time interval as labelled. Confidence contours indicate $99 \%$ (black), $90 \%$ (red), and $68 \%$ (green) confidence intervals for two interesting parameters. \label{fig:trao}}
\end{figure*}

For the absorption-dominated model, we carry out a similar systematic analysis as followed for the inner disk reflection model. We perform fits for all combinations of variable parameters for the following seven parameters: the column densities and covering fractions of both neutral absorbers, DRC normalization ($K_{DRC}$), PLC $\Gamma$, and the PLC $20-40 \keV$ flux ($F_{PLC}$). We allow parameters to be variable until we no longer find an improvement in the fit based on the F-test. We find a best-fit time-resolved fit with four variable parameters ($F_{PLC}$, $\Gamma$, and $N_H$ and $f_{cov}$ of PC2) as shown in Figure~\ref{fig:trao}. For the time-resolved analysis, we allow these four parameters to fit independently in each interval (which are shown in Table~\ref{tab:tsaao}) and allow all other free parameters (which are shown in Table~\ref{tab:tsbao}) to fit freely but linked between each interval.

We find a best-fit absorption-dominated model that provides a slightly worse fit of the time-resolved data ($\Delta\chi^2/\Delta\nu = +29/+5$) relative to the inner disk reflection model. Within the context of the absorption-dominated model, variation in the covering fraction of PC2 and the PLC flux drives the variability during the observations. The PC2 $N_H$ and $f_{cov}$ parameters appear to be correlated with variations in the coronal continuum on the short timescales seen in the top panel of Figure~\ref{fig:trao}. We show correlation plots in the bottom panel of Figure~\ref{fig:trao} to show that $f_{cov}$ and  $F_{PLC}$ have closed confidence contours.

\begin{deluxetable*}{ccccc}
\tablecaption{Time-Resolved Best-Fit Values for Parameters Free to Fit within Each Interval for the Absorption-Dominated Model  \label{tab:tsaao}}
\tablehead{\colhead{Int.} & \colhead{$f_{cov}$}  &  \colhead{$N_H$} & \colhead{$\mathrm{log}(F_{PLC}/\ergpcmsqps)$}  & \colhead{$\Gamma$} \\ 
\colhead{} &  \colhead{}  & \colhead{($10^{22} \pcmsq$)} & \colhead{}  & \colhead{}}
\startdata
1 & $0.30\pm0.08$ & $55\pm16$ & $-9.86_{-0.02}^{+0.03}$ & $1.94_{-0.06}^{+0.07}$ \\
2 & $0.35\pm0.03$ & $64\pm10$ & $-9.78_{-0.02}^{+0.01}$ & $1.95\pm0.03$ \\
3 & $0.49_{-0.04}^{+0.03}$ & $61_{-7}^{+6}$ & $-9.78_{-0.02}^{+0.01}$ & $1.93\pm0.04$ \\
4 & $0.51_{-0.03}^{+0.02}$ & $57_{-4}^{+5}$  & $-9.81_{-0.02}^{+0.01}$ & $1.91_{-0.03}^{+0.02}$ \\
5 & $0.42_{-0.05}^{+0.04}$ & $40\pm5$ & $-9.74_{-0.02}^{+0.01}$ & $1.83\pm0.03$ \\
6 & $0.32_{-0.08}^{+0.07}$ & $47\pm13$ & $-9.73_{-0.02}^{+0.01}$ & $1.85_{-0.06}^{+0.05}$ \\
7 & $0.41\pm0.05$ & $61_{-8}^{+7}$ & $-9.81_{-0.01}^{+0.02}$ & $1.91\pm0.04$ \\
Fit & $\chi^2/\nu$ & $15684/15594(1.01)$ &  &  \\
\tablecomments{$90 \%$ confidence intervals are given for the variable parameters for the absorption-dominated model, which are the PC2 covering fraction, PC2 column density, power-law component flux in the $20-40 \keV$ band, and power-law component photon index.}
\enddata
\end{deluxetable*}

\begin{deluxetable}{ccc}
\tablecaption{Time-Resolved Best-Fit Values for Parameters Assumed to not Vary During the Observations for the Absorption-Dominated Model \label{tab:tsbao}}
\tablehead{\colhead{Comp.} & \colhead{Par. (units)} & \colhead{Values}\\ \colhead{ } & \colhead{ } & \colhead{ }}
\startdata
PC1 &  $N_H$ ($\pcmsq$) & $13.5_{-1.0}^{+0.9}\times10^{22}$ \\
  & $f_{cov}$ & $0.94\pm0.01$ \\
WA1& $N_H$ ($\pcmsq$) & $(1.6\pm0.5)\times10^{22}$ \\
 & $\mathrm{log(}\xi/\ergcmps$)  & $2.5(f)$  \\
WA2& $N_H$ ($\pcmsq$) & $5.9_{-0.7}^{+0.9}\times10^{22}$ \\
 & $\mathrm{log(}\xi/\ergcmps$)  & $3.5\pm0.1$ \\
DRC & $\mathrm{A_{Fe}}$\tablenotemark{a} & $1.7_{-0.3}^{+0.2}$ \\
 & $K_{DRC}$ ($\phpcmsqps$) & $(4.1\pm0.3)\times 10^{-4}$ \\
 cross-norm. \tablenotemark{b} & XIS-1 & $0.95\pm0.01$ \\
 & FPMA& $1.03\pm0.01$  \\
 & FPMB & $1.06\pm0.01$  \\
 Fit & $\chi^2/\nu$ & $15684/15594(1.01)$  \\
\enddata
\tablecomments{$90 \%$ confidence intervals are given. WA1 and WA2 indicate warm absorbers one and two, respectively, PC1 indicates the partial covering, neutral absorber, DRC indicates the distant reflection component, and PLC indicates the power-law component.} 
\tablenotetext{a}{The Fe abundances of the IDR and DRC components are linked together in the fitting.}
\tablenotetext{b}{Cross-normalizations are relative to XIS-FI.}
\end{deluxetable}

\section{Discussion}\label{sec:d}

Using timing and spectral analyses, we have characterized the constant and variable spectral behavior of NGC 4151 as observed simultaneously with \textit{NuSTAR} and \textit{Suzaku}. Through a spectral analysis, we apply inner accretion disk reflection and absorption-dominated models. We find that the inner disk reflection model provides a better fit to the time-averaged and time-resolved spectra. Within the context of the inner disk reflection model, we find evidence for relativistic reflection from the inner accretion disk and a preliminary spin measurement for the supermassive black hole.

It is unlikely that a putative neutral absorber (which are at radii of $r\sim10^4-10^5 \rg$; \citealt{Antonucci:1993aa}, \citealt{Urry:1995aa}) can partially cover AGN coronae (which have been found to have size $D\lesssim10\rg$ for several AGNs including NGC 4151; \citealt{Reis:2013aa}). Furthermore, there is no physical reason to expect a neutral absorber to be correlated with variations in the coronal continuum on the short timescales seen in \S\ref{sec:traao} because the neutral absorbing gas likely is located relatively far from continuum source. A reasonable range of distances for the absorbing material might extend as close as the broad-line region ($r=6.6_{-0.8}^{+1.1}$ light-days; \citealt{Bentz:2006aa}) and as far as the upper-limit on the size of a dusty torus in the source ($r\sim 4.2\times10^4$ light-days; \citealt{Radomski:2003aa}). In contrast, a correlation between the inner disk reflection and the coronal continuum is physically possible based on light crossing time arguments and has been observed. The distance between the corona and the disk inferred by \citet{Cackett:2014aa} and the size scale of the inner disk, which are both $\lesssim 10 \rg$, correspond to a light-travel time $\lesssim 2 \ks$ assuming the black hole mass $M=4.57_{-0.47}^{+0.57} \times 10^{7} \Msun$ \citep{Bentz:2006aa}. The reverberation analysis of \citet{Zoghbi:2012aa} shows that the inner disk reflection flux can lag behind the coronal flux on time-scales $\sim 2 \ks$.  

Thus, on statistical and physical grounds, we discuss only the inner disk reflection model. Here, we discuss assumptions that may introduce systematic errors in the preliminary black hole spin value and physical scenarios for the disk-corona system. For calculations that use the black hole mass, we use $M=4.57_{-0.47}^{+0.57} \times 10^{7} \Msun$ \citep{Bentz:2006aa} measured from optical and UV reverberation.

\subsection{Systematic Uncertainties in the Black Hole Spin}\label{sec:d1}

Assumptions in our model introduce systematic uncertainty into the derived confidence intervals in the dimensionless black hole spin. A primary assumption of modeling the inner disk reflection with the {\tt relconv} kernel is that the accretion disk is radiatively efficient, optically thick, and geometrically thin at radii greater than or equal to the ISCO radius. These assumptions generally should be applicable to NGC 4151 considering the modest accretion rate implied by its bolometric luminosity $L_{bol}\sim 0.01L_{Edd}$. The assumption of a geometrically thin disk breaks down where the disk thickness, $t$, is on the same order as the radius in the disk, $r$, or $t/r \sim 1$. For Seyfert galaxies with Eddington ratios in the range $L_{bol}/L_{Edd} \sim 0.01-0.3$, we expect disk thicknesses in the range $t\sim (0.1 - 1.0) \rg$ for a \citet{Shakura:1973aa} disk. Thus, this assumption may be invalid for $r_{in} \lesssim 2 \rg$, which corresponds to spins $a\gtrsim 0.94$. Thus, considering the finite thickness of the disk close to the black hole, we estimate that this assumption relaxes our spin constraint to $a > 0.94$.  

We assume that the inner accretion disk radius is the ISCO radius and that a negligible level of emission originates within that radius. The latter assumption is supported by the near-maximal spin and relatively low Eddington ratio of the source, which may indicate that some of the accretion flow within the ISCO radius is optically thin \citep{Reynolds:2008aa}. Additionally, when we allow the inner edge of the accretion disk to fit freely, it fits to a value consistent with the ISCO radius although we restrict it to fit to values $r_{in}\geq r_{ISCO}$. Based on MHD simulations  \citep{Reynolds:2008aa}, if significant emission comes from within the ISCO radius, the fit will be biased to higher spins systematically with error $\lesssim 2 \%$ for rapidly spinning, prograde black holes ($a\geq 0.9$).   Thus, we estimate this assumption adds $\sim 2 \%$ systematic uncertainty to our spin value. Including the systematic uncertainties introduced by our assumptions about the thickness and extent of the inner disk, we find a constraint $a>0.9$. 

Black hole spin has some degeneracy with the iron abundance of the inner disk, as both properties affect the strength of the broad Fe line (e.g., \citealt{Reynolds:2012ab}). We find that NGC 4151 has a super-solar Fe abundance. There is no apparent degeneracy between our derived Fe abundance and spin values as shown in Figure~\ref{fig:contoursCor}. Fe abundance is constant on long time-scales. There is evidence for super-solar Fe abundance in material distant from the SMBH. From several \textit{Ginga} measurements, \cite{Yaqoob:1993aa} report an average value $\AFe\sim2.5$ from modeling an Fe K absorption edge. From analyzing the strength of a neutral Fe absorption edge in \textit{XMM-Newton} observations, \citet{Schurch:2003aa} report an Fe abundance $\AFe\sim3.6$, and they report $\AFe\sim2.0$ if they associate the Fe abundance of the neutral absorber with that of a Compton reflection component.  

For the Fe abundance of the inner disk, \citet{Nandra:2007aa} measure an Fe abundance in the range $\AFe=0.3-0.5$ from analyzing three \textit{XMM-Newton} observations. They assume the distant reflection component and inner disk reflection component (which are  modeled with \texttt{pexmon} and \texttt{kdblur2*pexmon}, respectively) are chemically homogeneous in the same way as in the present work. \citet{Nandra:2007aa} caution that their \textit{XMM-Newton} values require high sensitivity, broad-band observations that can simultaneously constrain the Fe K$\alpha$ line and Compton hump reflection features. Our analysis simultaneously constrains both features, and our derived Fe abundance ($\AFe=5.0_{-0.1}^{+0.0p}$) is moderately high in comparison with measurements of super-solar Fe abundance in material distant from the SMBH. This may introduce some bias in our spin constraint as a result of the positive correlation between spin and the Fe abundance \citep{Reynolds:2012ab}. However, we conclude that our derived Fe abundance negligibly biases our spin constraint of $a > 0.9$.

Black hole spin can show a degeneracy with the disk inclination if the blue wing of the broad Fe line profile is not significantly detected, as both parameters affect the width of the broad Fe line profile. A degeneracy is seen between the disk inclination and spin shown in Figure~\ref{fig:contoursCor}. The inclination of the inner disk should be constant on long time-scales, so we compare our determined inclination angle with various inclination angles previously reported for the source.

The inner disk inclination angle we find is consistent with most values determined for the inner disk. From the analysis of three \textit{XMM-Newton} observations,  \citet{Nandra:2007aa} find an inclination of the inner disk of $i=17_{-17}^{+12} \deg$, $i=21_{-21}^{+69} \deg$, and $i=33_{-3}^{+1} \deg$, respectively, where values are given for the best-fit \texttt{kdblur2*pexmon} model presented for each observation.  From an \emph{ASCA} observation, \citet{Yaqoob:1995aa} find $i=0_{-0}^{+19} \deg$ using the \texttt{diskline} model \citep{Fabian:1989aa}. \citet{Cackett:2014aa} report $i<30\deg$ for the inner disk from  Fe K$\alpha$ reverberation measurements.  

Comparing the inner disk inclination angle to that of more distant structures, our determined inner disk inclination angle is not consistent with the inclination of the narrow-line region, $45 \pm 5 \deg$, as determined from \textit{Hubble Space Telescope}  observations \citep{Das:2005aa}. This suggests that the inner disk and narrow-line region are misaligned. The inclination we find is consistent with the galactic disk inclination, $i\sim21\deg$, as inferred from optical photometry  \citep{Simkin:1975aa} and Very Large Array observations of neutral hydrogen \citep{Pedlar:1992aa}. 

It is possible that some of the broad Fe K line profile may arise from an alternate mechanism than reflection from the inner accretion disk. For example, scattering in a Compton thick wind near the disk can produce broadened Fe line profiles \citep{Sim:2008aa}.  However, the strength of this feature significantly depends upon the outflow mass loss rate. A low mass outflow rate is expected for the low mass accretion rate of the source implied by its Eddington ratio $L_{bol}/L_{Edd}\sim0.01$. Thus, we estimate that any contribution of scattering in a Compton thick wind near the accretion disk to the broad Fe line is relatively weak compared to the broad line profile from inner accretion disk reflection. 

\subsection{Physical Scenarios for the Corona and Inner Accretion Disk}\label{sec:d2}

Our derived high inner disk emissivity index ($q_1=10.0_{-0.4}^{+0.0p}$) indicates that the corona that illuminates the inner disk is compact. The inner emissivity index and photon index, $\Gamma=1.75_{-0.02}^{+0.01}$, together indicate a lamp post height $h<10 \rg$ assuming a lamp post geometry, as the maximum inner index for $\Gamma\sim1.8$, $h=10 \rg$, and $a=0.99$ is $q_1\sim3$ \citep{Dauser:2013aa}. A lamp post height $h\lesssim 10 \rg$ is expected in order to produce strong reflection features from which black hole spin can be well constrained (\citealt{Dauser:2013aa}; \citealt{Fabian:2014aa}). Furthermore, the high ionization parameter we find for the inner disk atmosphere is expected for a close illuminating source. We note that a high emissivity index ($q_1\gtrsim3$) is only physically consistent with a near-maximal black hole spin. 

We calculate theoretical emissivity profiles for the lamp post model of \citet{Dauser:2013aa}  for our best-fit $\Gamma$ and $a$ values. For this calculation, we assume a minimum height in order to give a rough estimate of the highest emissivity index expected. We estimate a value of $q_1\sim7$ if we average the emissivity from $r_{ISCO}$ to the emissivity break radius that we find for the best-fit model ($r_{br}$). So, we find an inner emissivity that is steeper than expected for the lamp post model of \citet{Dauser:2013aa}. This partially illustrates why the broken power-law emissivity profile for the inner disk provides a better fit to the data than the lamp post emissivity profile. However, this estimate does not necessarily indicate that our best-fit model is unphysical. 

We find a relatively low reflection fraction compared to the maximum expected reflection fraction, which is $R\sim 5-10$ for the lamp post model described in \S\ref{sec:taaidr} with $a=0.98\pm0.01$ and $h=1.3_{-0.0p}^{+0.1} \rg$ \citep{Dauser:2014aa}. This value is a factor $\sim4-8$ higher than what we find. There are several possible physical scenarios for the corona and inner accretion disk that may cause this apparent discrepancy. 

If the corona results from a mildly relativistic outflow rather than a static region, then the reflection fraction is expected to be reduced because of relativistic aberration (e.g., \citealt{Beloborodov:1999aa}, \citealt{Malzac:2001aa}). A photon index  $\Gamma\sim1.75$ indicates $v/c\sim 0.4$ near the base of an AGN jet and a reduction in the strength of reflection, $\Omega/2\pi$, by a factor of $\sim10$ using the disk inclination $i=18 \deg$ \citep{Beloborodov:1999aa}. This can help explain the apparently low reflection fraction we find. We note that VLBI observations have shown that the radio jet in NGC 4151 has had a component with velocity $v\leq0.050c$ at a distance $r\sim0.16 \pc$ from the nucleus \citep{Ulvestad:2005aa}. Thus, if the corona corresponds to the base of a mildly relativistic jet, then a mechanism is required to decelerate the jet to non-relativistic speeds on $\sim0.1\pc$ scales. Also, an outflow tends to flatten the emissivity profile \citep{Dauser:2013aa}. Keeping these facts in mind, an outflowing corona may help explain the low reflection fraction we find.

Truncation of the inner edge of the accretion disk at radius $r>r_{ISCO}$ can similarly reduce the expected reflection fraction.  \citet{Lubinski:2010aa} conclude that the relatively weak strength of reflection from the disk they measure ($\Omega/2\pi\simeq0.3$) from all \textit{INTEGRAL} observations of NGC 4151 from 2003 January to 2009 June can be explained by an inner hot accretion flow surrounded by a cold disk truncated at $r_{in}\sim15\rg$ or, alternatively, a mildly relativistic coronal outflow. When we allow the inner radius of the accretion disk to fit freely using the \texttt{kdblur3} model as described previously, it fits to a value consistent with the ISCO radius of a black hole with the near-maximal, prograde spin that we find for the best fit model. Thus, we do not find evidence for a truncated disk from spectral fitting of our observations. 

Dilution of the reflection features may also reduce the measured reflection fraction. For a close disk-illuminating source suggested by the steep emissivity profile, it is plausible that the inner disk sees an intense irradiating flux and has a tenuous, hot atmosphere. This may have a density profile roughly similar to that of an accretion disk in hydrostatic equilibrium, for which an intense irradiating flux can lead to disk thermal instability \citep{Nayakshin:2001ab, Ballantyne:2001ab}. This instability can produce a highly ionized `skin', which could both further broaden the disk reflection features and reduce the observed reflection fraction.  \citet{Ballantyne:2001ab} showed that simulated \textit{RXTE} and \text{ASCA} data of hydrostatic reflection models \citep[e.g.,][]{Ballantyne:2001ab} fit with constant density models (\citealt{Magdziarz:1995aa}; \citealt{Ross:1993aa}) tend to fit with systematically low reflection fraction values because of dilution. Although \textit{NuSTAR}  provides superior sensitivity in the $20-40 \keV$ energy band, where the reflection spectrum is almost completely independent of ionization parameter, our fit is biased towards best matching the spectral shape at lower energies ($E\lesssim20 \keV$), where the reflection spectrum is sensitive to the ionization parameter. 

With the \texttt{xillver} models, we assume that the disk is illuminated at $45 \deg$. However, for a compact coronal geometry, the disk illumination may be incident on the disk at closer to grazing angles, especially within a few $\rg$ of the black hole \citep{Dauser:2013aa}. In the innermost regions of the disk, this would enhance the influence of dilution and thus further reduce the reflection fraction. 

In summary, both dilution from a highly ionized disk skin and a mildly relativistic outflow may contribute to the reduced reflection fraction that we find relative to predictions for the lamp post geometry. Because an outflow tends to flatten the emissivity profile,  the formation of a highly ionized disk `skin' may be the dominant cause of the reduced reflection fraction. We do not find evidence for truncation of the inner accretion disk.

\subsection{Coronal Properties}\label{sec:d3}

We model the coronal emission with a cut-off power-law with $E_{Cut}=1000 \keV$ fixed and find $\Gamma=1.75_{-0.02}^{+0.01}$. This value is consistent within $3\sigma$ with the average Seyfert photon index, $\Gamma=1.84 \pm 0.03$, measured for 105 Seyfert galaxies with redshift $z\leq0.1$ with \textit{Bepposax} \citep{Dadina:2008aa}. It is relatively hard compared to the average photon index, $\Gamma=1.93\pm0.01$, for 144 Seyfert galaxies in the second \textit{INTEGRAL} AGN catalogue \citep{Beckmann:2009aa}. The value we find agrees within $1\sigma$ with the correlation between the photon index and Eddington ratio for a sample of 69 radio-quiet AGNs out to $z\sim2$ \citep{Brightman:2013aa}, which gives $\Gamma=1.63\pm0.16$ for $L_{bol}/L_{Edd}=0.01$. We note that the significant dispersion in this correlation as well as uncertainty in estimating the Eddington ratio should be noted when applied to a single source. 

The photon index we find is consistent with the range in photon index ($\Gamma\sim1.7-1.86$) that \citet{Lubinski:2010aa} measure for the source in a large range of X-ray flux states as observed with \emph{INTEGRAL} during the period 2003 January to 2009 June. The energy cut-off we find is corroborated by the energy cut-offs found by \citet{Lubinski:2010aa} for the source in high and medium flux states  ($E_{Cut}=264_{-26}^{+48} \keV$ and $E_{Cut}>1025 \keV$, respectively), which serve as upper and lower bounds, respectively, to the $20-100 \keV$ flux of the source in our data. 

\section{Conclusions}\label{sec:c}

We present X-ray timing and spectral analyses of contemporaneous, 150 $\ks$ \textit{NuSTAR} and \textit{Suzaku} X-ray observations of the Seyfert 1.5 galaxy NGC 4151. We test separate inner disk reflection and absorption-dominated models in order to determine the properties of the innermost regions of the AGN. Our primary conclusions are enumerated below:

1. We find that the inner disk reflection model provides a better statistical and physical description of the data than the absorption-dominated model for both the time-averaged and time-resolved spectral analyses.

2. Within the context of this inner disk reflection model, we find that relativistic emission originates in a highly ionized inner accretion disk with a steep inner emissivity profile, which suggests a bright, compact inner disk-illuminating source. We find a relatively moderate reflection fraction with respect to predictions for the lamp post geometry.  We find a preliminary, near-maximal black hole spin $a > 0.9$ accounting for statistical and systematic modeling errors.

3. Through a time-resolved spectral analysis, we find that modest coronal and inner disk reflection variation drive the spectral variability during the observations. 

4. We discuss physical scenarios with respect to our derived inner accretion disk and coronal properties. We find that a compact coronal geometry can reproduce the observed features.

This work further displays the ability of \textit{NuSTAR} in conjunction with X-ray observatories such as \textit{Suzaku},  \textit{XMM-Newton},  \textit{Chandra}, and  \textit{Swift} to robustly constrain coronal and accretion disk properties in Seyfert galaxies. 

\acknowledgments

This work was supported under NASA Contract No. NNG08FD60C, and made use of data from the \textit{NuSTAR} mission, a project led by the California Institute of Technology, managed by the Jet Propulsion Laboratory, and funded by the National Aeronautics and Space Administration. This research made use of the \textit{NuSTAR} Data Analysis Software (NuSTARDAS) jointly developed by the ASI Science Data Center (ASDC, Italy) and the California Institute of Technology (Caltech, USA). MLK gratefully acknowledges support through NASA grant \#NNX13AE90G. GM and AM acknowledge financial support from Italian Space Agency under grant ASI/INAF I/037/12/0-011/13 and from the European Union Seventh Framework Programme (FP7/2007-2013) under grant agreement n.312789. CSR acknowledges support from the NASA-ADAP program under grants NNX14AF86G and NNX14AF89G. We thank Alan Marscher for helpful discussions. We thank the anonymous referee for their comments, which have improved this manuscript. This research made use of Astropy, a community-developed core Python package for Astronomy \citep{Astropy-Collaboration:2013aa}, and Matplotlib \citep{Hunter:2007aa}. 

\textit{Facilities:} \facility{NuSTAR (FPMA, FPMB)}, \facility{Suzaku (XIS, HXD)}

\appendix  

\section{Influence of the Minor Pileup in Suzaku XIS data} \label{sec:appA}

To investigate the influence of the mild pileup in the XIS data on our results, we extract XIS data from an annular region with inner radius $30 \arcsec$ and outer radius $170 \arcsec$. We fit our best-fit model to this data, and we find consistent parameter values within $90\%$ confidence in one interesting parameter except for instrument cross-normalizations relative to XIS-FI. We find cross-normalizations  for XIS-1 of $0.93\pm0.01$, PIN of $1.21 \pm 0.01$,  FPMA of $1.00\pm0.01$, and FPMB of  $1.03\pm0.01$. Thus, we find that the minor pileup fraction does not influence our conclusions on the spectral properties of NGC 4151.

\bibliography{refs}

\end{document}